\documentclass{aa} 
\pdfoutput=1 

\usepackage{graphicx}
\usepackage{txfonts}

\begin{document}

   \title{Primordial black holes in the main lensing galaxy of FBQ 0951+2635}    
                                                             
   \author{Daniel Isla\inst{1}
		\and
	  	Luis J. Goicoechea\inst{1,2}
		\and
		Ana Esteban-Guti\'errez\inst{3}
		\and
		Vyacheslav N. Shalyapin\inst{1,2,4}
		\and
		Rodrigo Gil-Merino\inst{5}
		\and
                Jose M. Diego\inst{2}
                \and
                Eleana Ruiz-Hinojosa\inst{1}
                }

\institute{Departamento de F\'i sica Moderna, Universidad de Cantabria, 
                Avda. de Los Castros s/n, E-39005 Santander, Spain\\
		\email{disla003@ikasle.ehu.eus, goicol@unican.es}
                \and
                Instituto de F\'isica de Cantabria (CSIC-UC), Avda. de Los Castros s/n, 
		E-39005 Santander, Spain
		\and
		Instituto de F\'isica y Astronom\'ia, Universidad de Valpara\'iso, Avda. 
                Gran Breta\~na 1111, Valpara\'iso, Chile
                \and
 		O.Ya. Usikov Institute for Radiophysics and Electronics, National 
                Academy of Sciences of Ukraine, 12 Acad. Proscury St., UA-61085 
                Kharkiv, Ukraine
		\and
		Escuela Superior de Ingenier\'ia y Tecnolog\'ia, Universidad Internacional 
                de La Rioja (UNIR), Avda. Gran V\'ia Rey Juan Carlos I 41, E-26005 
                Logro\~{n}o, Spain
                }

 
\abstract{Although dark matter in galaxies may consist of elementary particles different 
from those that make up ordinary matter and that would be smoothly distributed (still 
undetected), the so-called primordial black holes (PBHs) formed soon after the initial Big 
Bang are also candidates to account for a certain fraction of mass in galaxies. In this 
paper, we focused on the main lensing galaxy ($z$ = 0.260) of the doubly imaged 
gravitationally lensed quasar FBQ 0951+2635 ($z$ = 1.246) for probing possible PBH 
populations. Assuming that the mass of the galaxy is due to smoothly distributed matter 
(SDM), stars, and PBHs, the 16-yr observed microlensing variability was compared in detail 
with simulated microlensing signals generated by 90 different physical scenarios. Among other 
details, the simulated signals were sampled as the observed one, and the observed variability 
in its entirety and over the long term were used separately for comparison. While none 
of the scenarios considered can reproduce the overall observed signal, the observed long-term 
variability favours a small mass fraction in PBHs with a mass of the order of the mean 
stellar mass. Furthermore, it is possible to obtain strong constraints on the galaxy mass 
fraction in Jupiter-mass PBHs, provided that a reverberation-based measurement of the source 
size is available and relatively small. To constrain the mass fraction in $\sim$10 
$\rm{M_{\odot}}$ PBHs, light curves five times longer are probably required.} 
   

\maketitle

\section{Introduction}
\label{sec:introd}

In the last quarter of the 20th century, several pioneering studies presented compelling 
evidence for the presence of significant amounts of dark matter in galaxy halos 
\citep[e.g.,][]{1978ApJ...225L.107R,1980ApJ...238..471R}. Despite all efforts to identify 
the composition of this invisible dark matter, its nature is still unknown. While tens of
elementary particles have been proposed as possible dark matter candidates 
\citep[e.g.,][]{2010Natur.468..389B}, primordial black holes (PBHs) formed in the early 
universe are non-luminous astrophysical objects that could also populate galaxy halos and be 
part of the dark matter on galaxy scales \citep[e.g.,][]{2021arXiv211002821C}. 

For the Milky Way and the Large Magellanic Cloud, strong constraints on the fraction of dark 
matter in planetary-mass and stellar-mass PBHs were derived through microlensing variability
of stars in the Milky Way's biggest satellite galaxy. A recent analysis has revealed that 
possible PBHs in the mass range 1.8$\times$10$^{-4}$--6.3 M$_{\odot}$ would make up 
$\leq$1\% of dark matter in these two galaxies, and that hypothetical populations of PBHs in 
the mass window 10$-$100 M$_{\odot}$ cannot compose more than 3\% of dark matter 
\citep{2024Natur.632..749M}. Therefore, gravitational microlensing observations indicate 
that planetary-mass and stellar-mass PBHs may only account for a small fraction of dark 
matter in both local galaxies. 

PBHs in non-local galaxies can also be detected from gravitational wave experiments (if they 
belong to coalescing binary systems) or from their gravitational microlensing effects on 
background compact sources. For example, the discovery of gravitational waves from non-local 
black-hole binaries by the LIGO-Virgo collaboration \citep[e.g.,][]{2019PhRvX...9c1040A} 
suggested the possibility that some of these black holes were formed in the early universe. 
For the 10 confidence detections of black-hole binaries in \citet{2019PhRvX...9c1040A}, the 
typical mass of black holes varied between 8 and 50 M$_{\odot}$, and they were located in 
the redshift interval 0.1 $\leq z \leq$ 0.5. However, the microlensing variability of a 
distant star ($z$ = 1.49) in the field of the galaxy cluster \object{MACSJ1149.5+2223} ($z$ 
= 0.54) favoured a small mass fraction of the intracluster medium in 
$\sim$30 $\rm{M_{\odot}}$ PBHs \citep[e.g.,][]{2018NatAs...2..334K,2018ApJ...857...25D}. 
Additionally, differential microlensing magnifications between pairs of images of lensed 
quasars indicated that the lensing galaxy mass in PBHs in the LIGO-Virgo mass window must be 
$\leq$1\% at the 90\% confidence level \citep{2022ApJ...929..123E}.

Microlensing in gravitationally lensed quasars is sensitive to possible populations of PBHs 
in lensing galaxies and has a great potential to constrain these populations 
\citep[e.g.,][and the previous paragraph]{2024SSRv..220...14V}. In this vein,
microlensing-induced energy shifts of the Fe K$\alpha$ emission line in three lensed quasars 
indicated that planetary-mass PBHs cannot compose more than 0.01$-$0.03\% of the total mass 
of the lensing galaxies at $z \sim$ 0.3 and 0.7 \citep{2018ApJ...853L..27D,
2019ApJ...885...77B}. However, single-epoch fluxes of samples of lensed quasars suggested 
that PBHs with substellar masses of 0.082 and 0.0024 M$_{\odot}$ could constitute up to 
$\sim$15\% and $\sim$60\% of the total mass at the 90\% confidence level, respectively 
\citep{2023ApJ...954..172E}. In addition, stars and star-like PBHs (those having a mass 
close to the mean stellar mass) can only amount to about 10$-$20\% of the total mass 
\citep[e.g.,][]{2009ApJ...706.1451M,2017ApJ...836L..18M}. 

Although observed microlensing variability of lensed quasars may also provide critical clues 
about the dark matter composition in lensing galaxy halos, recent studies have led to 
inconclusive or contradictory results. Using microlensing variations in six lensed quasars 
monitored for $\sim$10 years by the COSMOGRAIL collaboration, \citet{2023A&A...673A..88A} found 
that a standard scenario with only stellar microlenses (absence of PBHs) cannot be rejected. 
They also could not reject a scenario in which all dark matter in the galaxies at 0.3 $\leq 
z <$ 0.9 is due to stellar-mass PBHs. However, \citet{2020A&A...633A.107H} reached 
a very different conclusion, claiming that stellar-mass PBHs are required to explain 
amplitudes of microlensing signals in lensed quasars. Additionally, notice that these dark 
objects would be distributed either inside the lensing galaxies or along the lines of sight 
to the quasars, the latter possibility being supported by light curves and emission lines of 
non-lensed quasars \citep{2022MNRAS.512.5706H,2024MNRAS.527.2393H}. 

However, it is not a straightforward task to accurately simulate microlensing effects of a 
cosmologically distributed population of PBHs at different redshifts between a distant 
quasar and the observer. Also, the putative PBHs are expected to be concentrated at mass 
density peaks harbouring massive galaxies. Thus, we focus on a standard framework based on 
the presence of PBHs in main lensing galaxies of gravitationally lensed quasars. In this 
paper, we discuss the feasibility of different populations of PBHs in the main lensing 
galaxy of the doubly imaged quasar \object{FBQ 0951+2635} \citep{1998AJ....115.1371S} from a 
detailed analysis of its microlensing variability. In Sect.~\ref{sec:q0951}, we present 
lensing mass solutions and observed microlensing curves spanning 16 years (2008$-$2023). 
Sect.~\ref{sec:method} describes our methodology, which basically relies on the comparison 
between observed and synthetic (simulated) microlensing signals. Results are included in 
Sect.~\ref{sec:simres} and our conclusions are summarized in Sect.~\ref{sec:conclu}.

\section{FBQ 0951+2635: lensing mass solutions and observed difference light curves}
\label{sec:q0951}

We selected \object{FBQ 0951+2635} \citep{1998AJ....115.1371S} as a case study to learn 
about the potential of microlensing variability to constrain populations of PBHs 
in lensing galaxies. The brightest optical image of the double quasar is denoted by the 
letter A, while the faintest optical image and the main lensing galaxy are denoted by B and 
G, respectively. In a pioneering work on this lens system, \citet{2005A&A...431..103J} 
reported a quasar redshift of 1.246, a solution for the astrometry of ABG and the 
ellipticity of G (based on near-IR observations with the $Hubble$ Space 
Telescope)\footnote{These observations \citep{2000ApJ...543..131K} yielded other 
astro-photometric solutions \citep[e.g.,][]{2012A&A...538A..99S,2023ApJ...952...54R}, and 
here we do not consider different constraint sets through the same observations. 
However, the controversy on the galaxy light distribution is discussed in 
Sect.~\ref{sec:lensmass}}, a flux ratio of $B/A$ = 0.21 $\pm$ 0.03 at 8.4 GHz, and a time 
delay between both quasar images of 16 $\pm$ 2 d (A is leading). The 16-d delay relied on 
optical light curves from October 2000 to June 2001. Additionally, the redshift of G ($z$ 
= 0.260) was spectroscopically measured by \citet{2007A&A...465...51E}.

\subsection{Lensing mass solutions}
\label{sec:lensmass}
 
Using all the observational constraints of the system in the previous paragraph along with a 
flat $\Lambda$CDM cosmology with $H_0$ = 70 km s$^{-1}$ Mpc$^{-1}$, $\Omega_{\rm{M}}$ = 0.3, 
and $\Omega_{\Lambda}$ = 0.7, it is possible to model the lensing mass as a singular 
power-law ellipsoid (SPLE) describing the gravitational effects of G plus external shear 
(ES) due to intervening galaxies other than G. In this approach, the external convergence 
(EC or $\kappa_{\rm ext}$) is assumed to be negligible. The corresponding lensing mass 
solution \citep{MScThesis} allowed us to estimate the total convergence ($\kappa$), the 
total shear strength ($\gamma$), and the shear direction ($\theta_{\gamma}$) at the 
positions of A and B (see the second row in Table~\ref{tab:massol}). Unfortunately, 
regarding the astro-photometric solution of \citet{2005A&A...431..103J}, there is evidence 
that the effective radius of the light distribution of G is severely underestimated 
\citep{2012A&A...538A..99S,2025A&A...694A..31S}. In addition to this "size problem" of the 
elliptical light halo, \citet{2023ApJ...952...54R} suggested the existence of a disc, which 
was not considered in most previous solutions. Hence, due to the current ambiguity in the 
galaxy light distribution, we cannot use tight constraints on the convergence in stars.

To check the influence of the time delay/mass model on our final results, we also considered 
a second approach (see the third row in Table~\ref{tab:massol}). Since the 16-d delay from a 
short monitoring campaign has been questioned \citep[e.g.,][]{2011A&A...536A..44E,
2015A&A...580A..38R}, we used light curves from the Nordic Optical Telescope 
\citep[1999$-$2001;][]{2005A&A...431..103J,2006A&A...455L...1P}, the Kaj Strand Telescope 
\citep[2008$-$2017;][]{2023ApJ...952...54R}, and the Liverpool Telescope \citep[2009$-$2023; 
initial light curves were presented by][]{2018A&A...616A.118G} to measure a more robust 
delay of 13.3 $\pm$ 1.7 d. This time delay interval practically coincides with that of 
\citet{2025A&A...694A..31S}. In addition, \citet{2017ApJ...850...94W} conducted a 
spectroscopic survey of galaxies along the line-of-sight towards \object{FBQ 0951+2635}, 
showing evidence of an EC above 0.17. Taking into account the new constraint on the time 
delay, the lower limit $\kappa_{\rm ext}$ = 0.17, and the model Hubble constant $H_0/(1 - 
\kappa_{\rm ext})$, Ruiz-Hinojosa's mass solution (SPLE+ES model) is still usable as an 
"effective" solution. It is also easy to show that the total convergence is $\kappa = 
\kappa_{\rm G} + \kappa_{\rm ext}$, where $\kappa_{\rm G} = (1 - \kappa_{\rm ext}) \times 
\kappa_{\rm eff}$ is the actual convergence produced by G and $\kappa_{\rm eff}$ is the 
overestimated value of the SPLE+ES effective solution. Similarly, the total shear strength 
is $\gamma = (1 - \kappa_{\rm ext}) \times \gamma_{\rm eff}$, with $\gamma_{\rm eff}$ being 
the effective shear strength \citep[for definitions of the convergence and shear, and 
their effective or scaled versions, see, e.g.][]{1992grle.book.....S,1996ApJ...464...92G}.

\begin{table}[h!]
\begin{center}
\caption{Convergence and shear parameters for \object{FBQ 0951+2635}.}
\label{tab:massol}
\begin{tabular}{lcccccc}
   \hline \hline
   Approach & $\kappa_{\rm A}$ & $\gamma_{\rm A}$ & $\theta_{\gamma_{\rm A}}$ &
   $\kappa_{\rm B}$ & $\gamma_{\rm B}$ & $\theta_{\gamma_{\rm B}}$\\
   \hline 
   First   & 0.279 & 0.380 & 36.3 & 1.194 & 1.352 & 49.9\\
   Second  & 0.401 & 0.315 & 36.3 & 1.161 & 1.122 & 49.9\\
   \hline
\end{tabular}
\end{center}
\footnotesize{Note: In the first approach, we consider a 16-d time delay and a SPLE+ES mass
model, whereas the second approach relies on a shorter delay of 13.3 d and a SPLE+ES+EC mass 
model (see main text). The shear direction ($\theta_{\gamma}$) is given in degrees east of 
north.}
\end{table}

\subsection{Observed microlensing signals}
\label{sec:obsmicro}

Microlensing signals are derived from difference light curves, and the main idea is as 
follows. The flux of the quasar image B at time $t$ is given by $F_{\rm B}(t) = F_{\rm I}(t) 
\epsilon_{\rm B} \mu_{\rm B}(t)$, where $F_{\rm I}$ is the intrinsic quasar flux, 
$\epsilon_{\rm B}$ is the dust extinction factor, and $\mu_{\rm B}$ is the lens 
magnification. This lens magnification may vary over time as a result of microlensing 
effects. Similarly, after taking into account the time delay between 
the two quasar images ($\Delta t$), the flux of A verifies $F_{\rm A}(t-\Delta t) = F_{\rm 
I}(t) \epsilon_{\rm A} \mu_{\rm A}(t)$. To remove the intrinsic signal and convert fluxes to 
magnitudes, we compute the flux ratio, take its logarithm, and multiply by $-$2.5. The 
resulting magnitude difference can be written as 
\begin{equation}
\begin{split}
   B(t) - A(t-\Delta t) &= -2.5\log \left[ \frac{F_{\rm B}(t)}{F_{\rm A}(t-\Delta t)} 
   \right] \\
   &= -2.5\log \left[ \frac{\epsilon_{\rm B} \mu_{\rm B}(t)}{\epsilon_{\rm A} 
   \mu_{\rm A}(t)} \right] .
\end{split}
\label{eq1}
\end{equation}
By subtracting the 16-yr average magnitude difference from Eq.~\ref{eq1}, it is also 
possible to remove dust extinction effects. Therefore, we obtain
\begin{equation}
\begin{split}
   B(t) - A(t-\Delta t) &- \langle B(t) - A(t-\Delta t) \rangle \\
   &= -2.5\left\{ \log \left[ \frac{\mu_{\rm B}(t)}{\mu_{\rm A}(t)} \right] - \biggl 
   \langle \log \left[ \frac{\mu_{\rm B}(t)}{\mu_{\rm A}(t)} \right] \biggr \rangle 
   \right\}.
\end{split}
\label{eq2}
\end{equation}
Given the observed light curves $A(t)$ and $B(t)$, as well as a time delay measurement 
$\Delta t$, the expression on the left side of Eq.~\ref{eq2} allows us to build an observed 
difference light curve (ODLC). In the absence of microlensing effects, $\mu_{\rm 
B}(t)/\mu_{\rm A}(t)$ would remain constant and equal to the macro-magnification ratio, 
which means that the ODLC should be zero at all times.

\begin{figure}
\centering
\includegraphics[width=9cm]{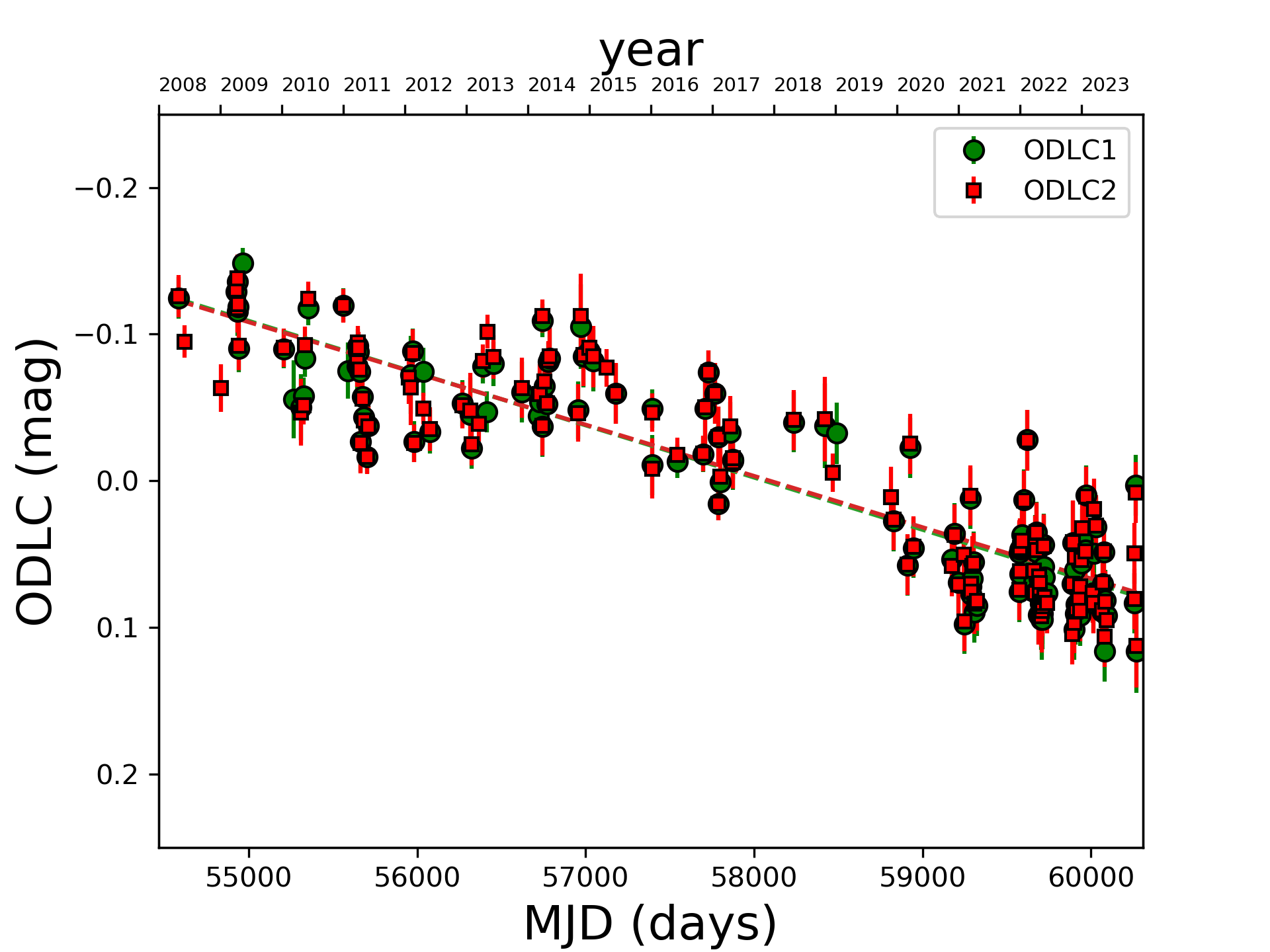}
\caption{Observed difference light curves of \object{FBQ 0951+2635} in the $r$ band. They 
are built from time delay values of 16 d (ODLC1) and 13.3 d (ODLC2). For each delay, the 
light curve of A is time shifted, and its shifted magnitudes around the dates in B are 
interpolated and subtracted from the magnitudes of B. The final step is to subtract the 
average magnitude difference. We also show two linear fits (green and red dashed lines) to 
the observed microlensing signals (altough they can barely be distinguished from each other 
in the plot).}
\label{fig:odlcs}
\end{figure}

We built two ODLCs using the two delay values corresponding to the approaches given in 
Table~\ref{tab:massol}. Regarding the quasar light curves, we used GLENDAMA+ brightness 
records of \object{FBQ 0951+2635} until 1 December 2023 \citep{2025A&A...694A..31S}, which 
were complemented with magnitudes at the two additional epochs MJD = 58925.941 and 
60266.264. These records mainly consist of $r$-band magnitudes provided by observations with 
the Liverpool Telescope \citep[GLENDAMA project;][]{2018A&A...616A.118G} and the Kaj Strand 
Telescope \citep{2023ApJ...952...54R}, and cover a period of 16 years in the last two 
decades (2008$-$2023). Despite the availability of earlier monitorings in the $R$ band 
\citep[e.g.,][]{2005A&A...431..103J,2006A&A...455L...1P,2009MNRAS.397.1982S}, we focused on 
well-sampled $r$-band light curves without significant gaps (see Sect.~\ref{sec:data}). The 
two ODLCs are displayed in Figure~\ref{fig:odlcs}. There is a very high degree of similarity 
between both observed microlensing signals (green circles and red squares in 
Figure~\ref{fig:odlcs}) since the correlation coefficient is 0.996.

The ODLCs of \object{FBQ 0951+2635} show a long-term variability (see the two linear 
fits in Figure~\ref{fig:odlcs}) along with rapid fluctuations around it. This short 
time-scale (rapid) variability could be due to standard microlensing of a compact source, 
observational noise or physics that is ignored when modelling in Sect.~\ref{sec:method}. 
There is growing evidence that both the compact accretion disc and the broad line region of 
some lensed quasars contribute significantly to their optical band fluxes 
\citep[e.g.,][]{2018A&A...616A.118G,2022A&A...659A..21P,2023A&A...677A..94F}, so that the 
rapid variability in the corresponding ODLCs may be correlated with intrinsic short-term 
variations \citep[see Fig. 4 of][]{2024MNRAS.530.2273G}. 
 
\section{Methodology}
\label{sec:method}

\subsection{Magnification maps}
\label{sec:magmaps}

We compared the ODLCs and synthetic difference light curves (SDLCs) from source trajectories 
on simulated magnification maps. These magnification maps for each quasar image were made 
using a Fortran-90 code based on the Poisson and Inverse Polygon (PIP) method (see 
Sect.~\ref{sec:data}), and a simpler 
version addressing a single population of microlenses is described in 
\citet{2021A&A...653A.121S}. The analysis was carried out according to two approaches 
(already detailed in Sect.~\ref{sec:q0951}): our first approach assumes that the convergence 
is exclusively due to the lensing galaxy G ($\kappa_{\rm ext} \sim$ 0), so it can be 
decomposed into three contributions \citep[e.g.,][]{2022ApJ...929..123E}: smoothly 
distributed matter (SDM) in the galaxy halo ($\kappa_{\rm sdmG}$), stars ($\kappa_{\rm 
starG}$), and PBHs ($\kappa_{\rm pbhG}$). Thus, in addition to the convergence and shear 
strength, the SDM mass fraction $f_{\rm sdmG} = \kappa_{\rm sdmG}/\kappa_{\rm G}$ and 
$F_{\rm pbhG} = \kappa_{\rm pbhG}/(\kappa_{\rm starG} + \kappa_{\rm pbhG})$ are other two 
parameters of the PIP software. Once $f_{\rm sdmG}$ and $F_{\rm pbhG}$ have been fixed, it 
is straightforward to know the mass fractions of the two microlens populations, i.e., 
$f_{\rm pbhG} = F_{\rm pbhG} (1 - f_{\rm sdmG})$ and $f_{\rm starG} = 1 - F_{\rm pbhG} (1 - 
f_{\rm sdmG}) - f_{\rm sdmG}$. In this case study, we considered a grid consisting of 
three relevant values of $f_{\rm sdmG}$: 0.1 (microlens dominated mass), 0.5, and 0.9 (SDM 
dominated mass), and four values of $F_{\rm pbhG}$: 0 (standard scenarios), 0.1, 0.5, and 
0.9. 

For the second approach, the convergence due to SDM includes both the contribution from G 
($\kappa_{\rm sdmG}$) and that from galaxy group halos acting as secondary deflectors 
($\kappa_{\rm ext}$). These groups are located near G and along the line of sight to the 
quasar \citep{2017ApJ...850...94W}. More specifically, $f_{\rm sdm} = \kappa_{\rm 
ext}/\kappa + f_{\rm sdmG} (1 - \kappa_{\rm ext}/\kappa)$, where $f_{\rm sdmG}$ is the SDM 
mass of G relative to its total mass (see above). We took again the 2D grid $\{[f_{\rm 
sdmG}], [F_{\rm pbhG}]\} = \{[0.1, 0.5, 0.9], [0, 0.1, 0.5, 0.9]\}$ to be consistent with 
the grid used in the first approach. 

With respect to the mass of the microlenses, stars are randomly distributed following a 
power-law mass function $dN/dM \propto M^{-\alpha}$ over a mass range $M_1 < M < M_2$, where 
$M_2/M_1$ denotes the maximum-to-minimum mass ratio. Some quasar microlensing studies relied 
on a Salpeter mass function ($\alpha$ = 2.35) with $M_2/M_1$ = 100 
\citep[e.g.,][]{2004ApJ...605...58K,2023A&A...673A..88A}, while others used a Kroupa mass 
function ($\alpha$ = 1.3) with $M_2/M_1$ = 50 \citep[e.g.,][]{2020ApJ...905....7C,
2023ApJ...952...54R}. However, all reasonable stellar mass functions produce similar 
microlensing effects \citep[e.g.,][]{2004ApJ...605...58K}, and we have taken a Kroupa 
distribution with $M_2/M_1$ = 50. We have also adopted a typical mean stellar mass $M_{\rm 
star}$ = 0.3 $\rm{M_{\odot}}$ \citep[e.g.,][]{2011ApJ...738...96M}. Due to our lack of 
knowledge about the properties of the other family of microlenses (PBHs), we assumed that 
PBHs are randomly distributed and have the same mass ($M_{\rm pbh}$), and considered three 
different values of $r_{\rm pbh} = \log (M_{\rm pbh}/M_{\rm star})$. These values are 
$r_{\rm pbh}$ = $-$2.5 ($M_{\rm pbh} \sim$ 0.001 $\rm{M_{\odot}}$; Jupiter-mass PBHs), 
$-$0.5 ($M_{\rm pbh} \sim$ 0.1 $\rm{M_{\odot}}$), and 1.5 ($M_{\rm pbh} \sim$ 10 
$\rm{M_{\odot}}$; PBHs with stellar black hole mass). 

\begin{figure*}
\centering
\includegraphics[width=7.5cm]{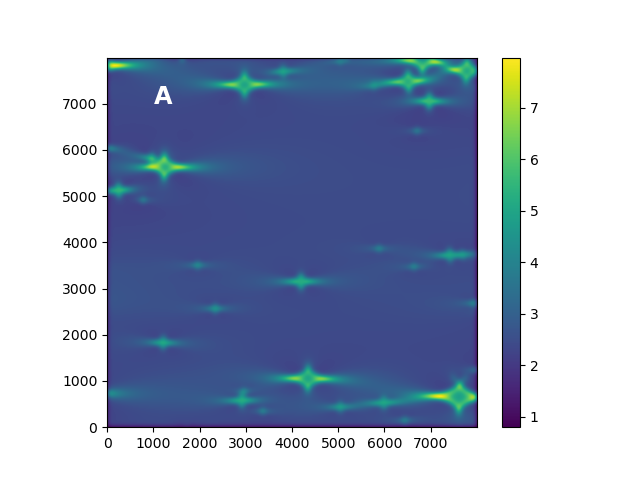}
\includegraphics[width=7.5cm]{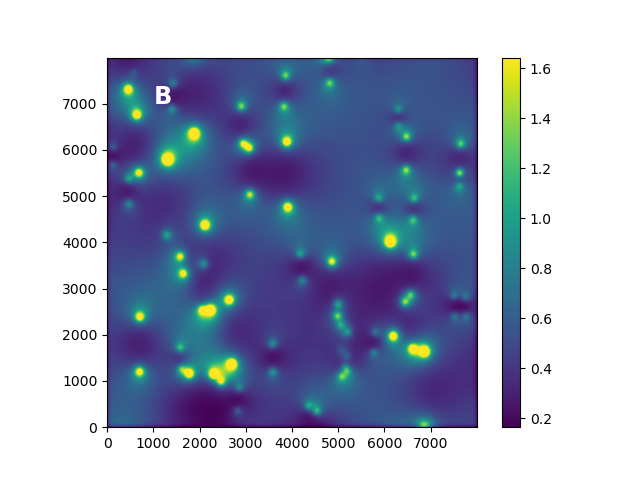}
\includegraphics[width=7.5cm]{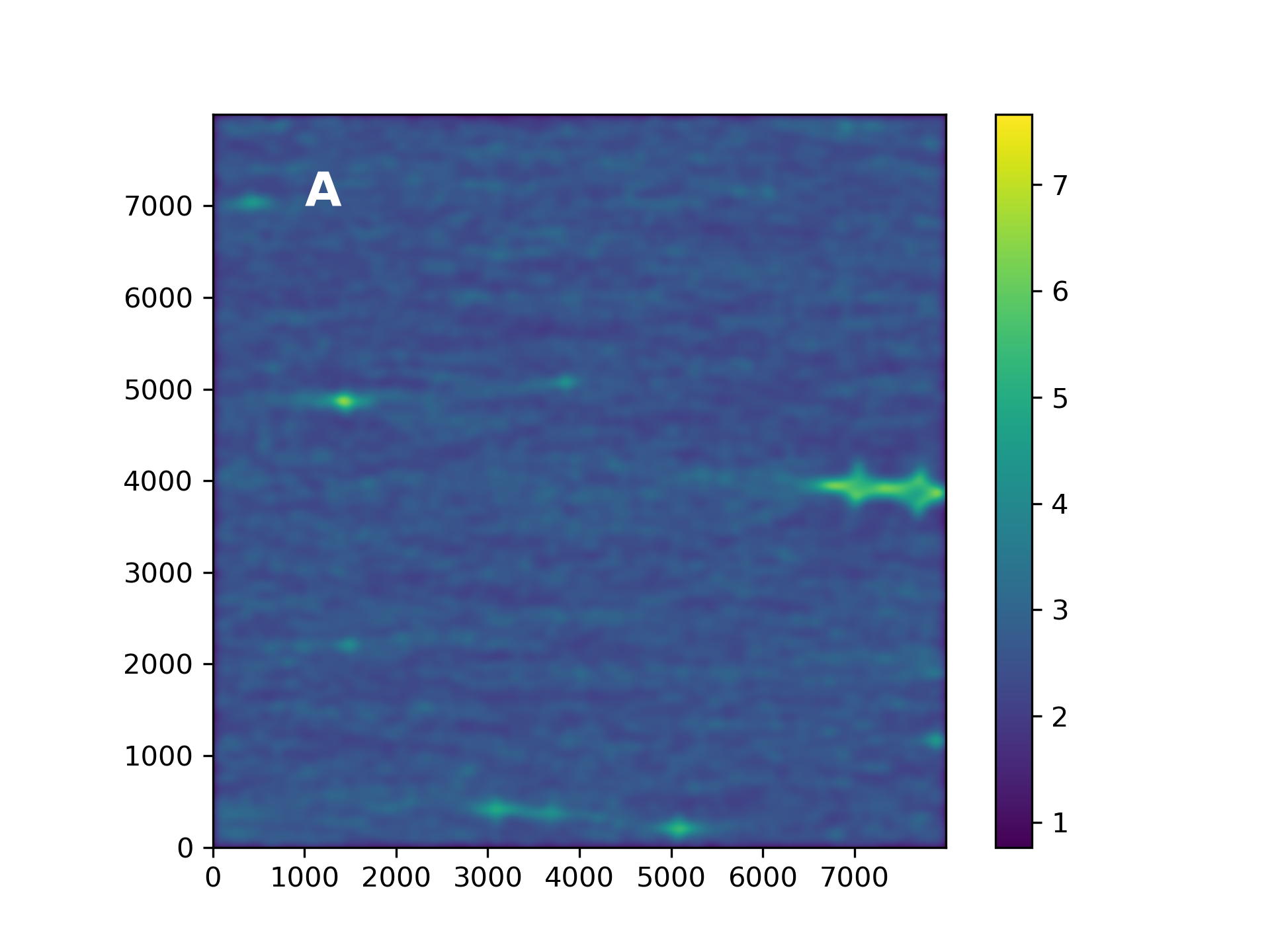}
\includegraphics[width=7.5cm]{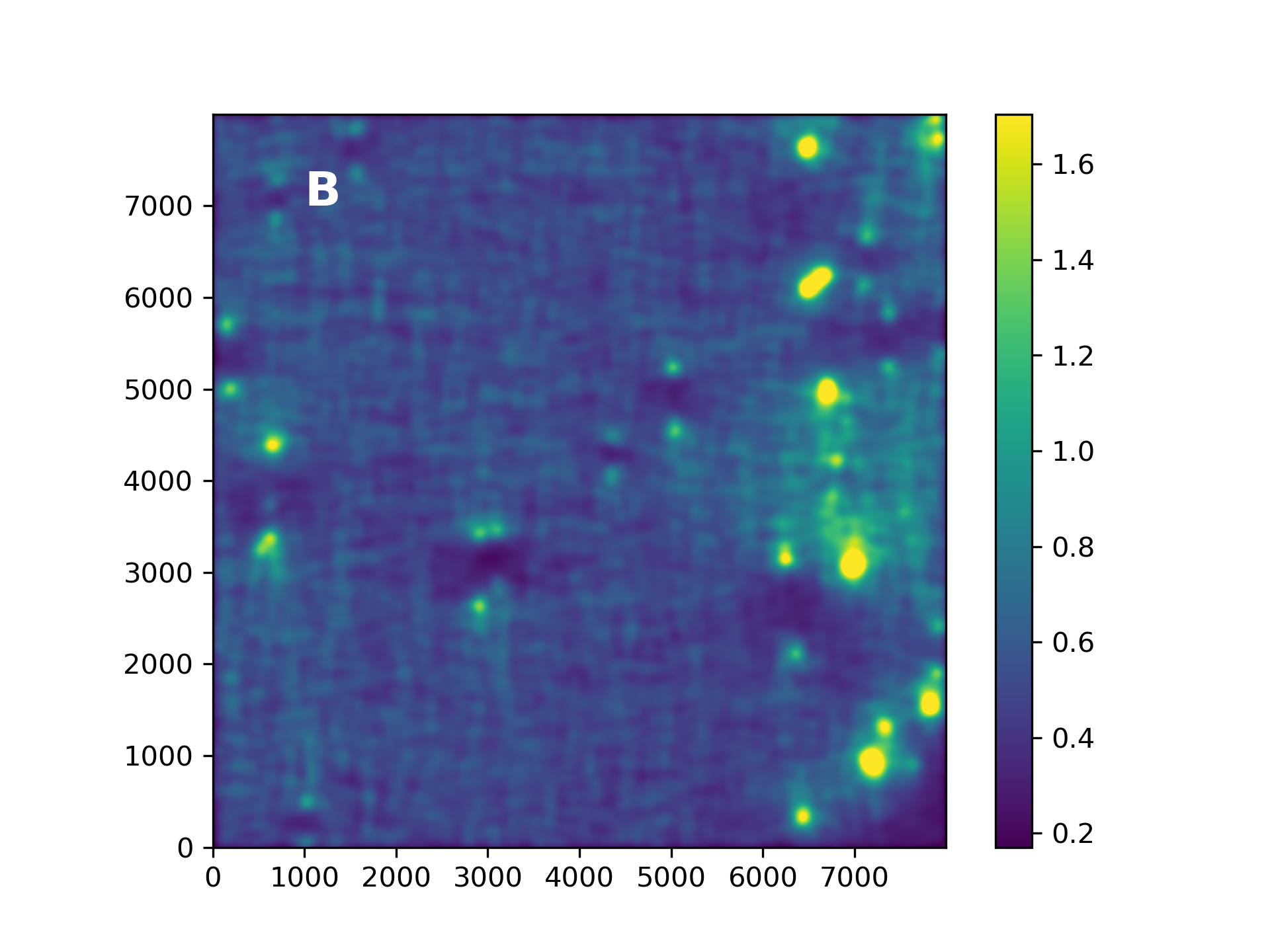}
\includegraphics[width=7.5cm]{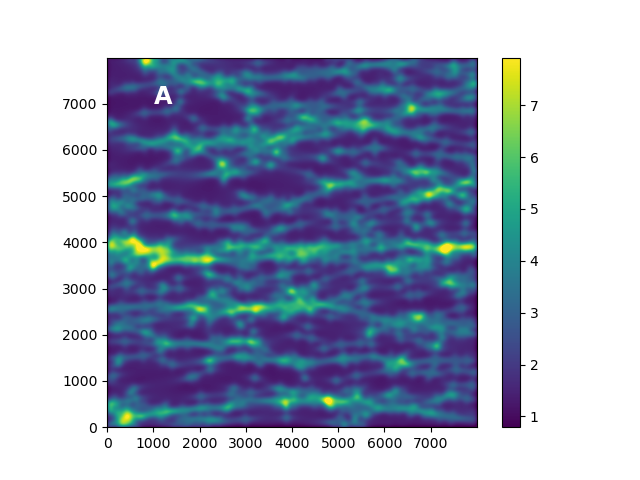}
\includegraphics[width=7.5cm]{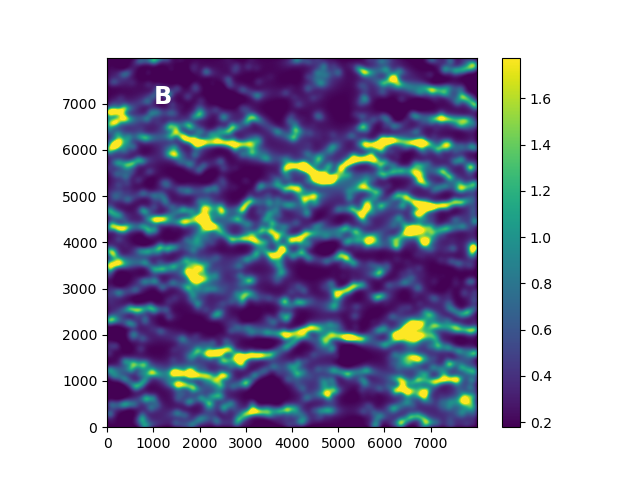}
\includegraphics[width=7.5cm]{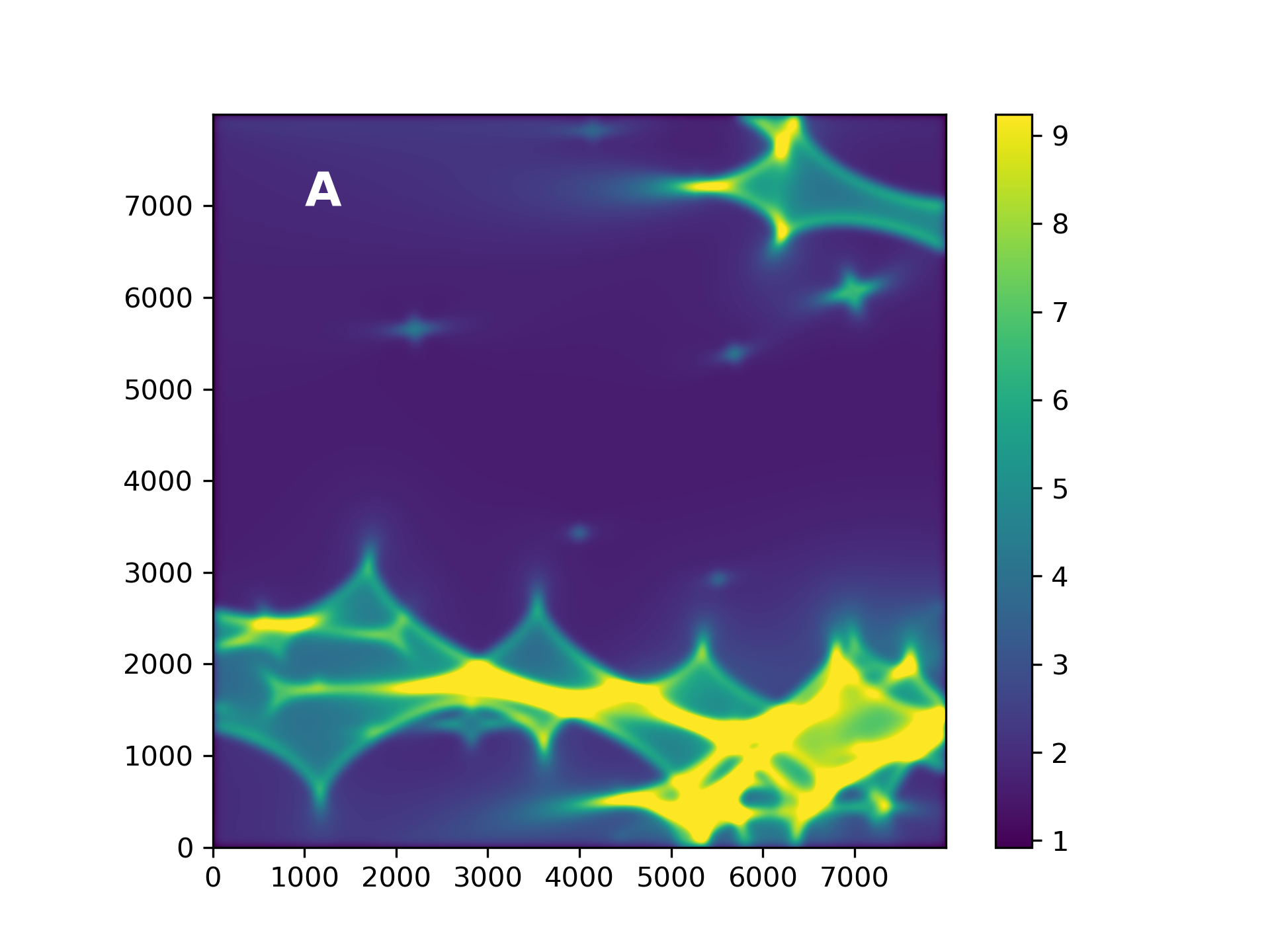}
\includegraphics[width=7.5cm]{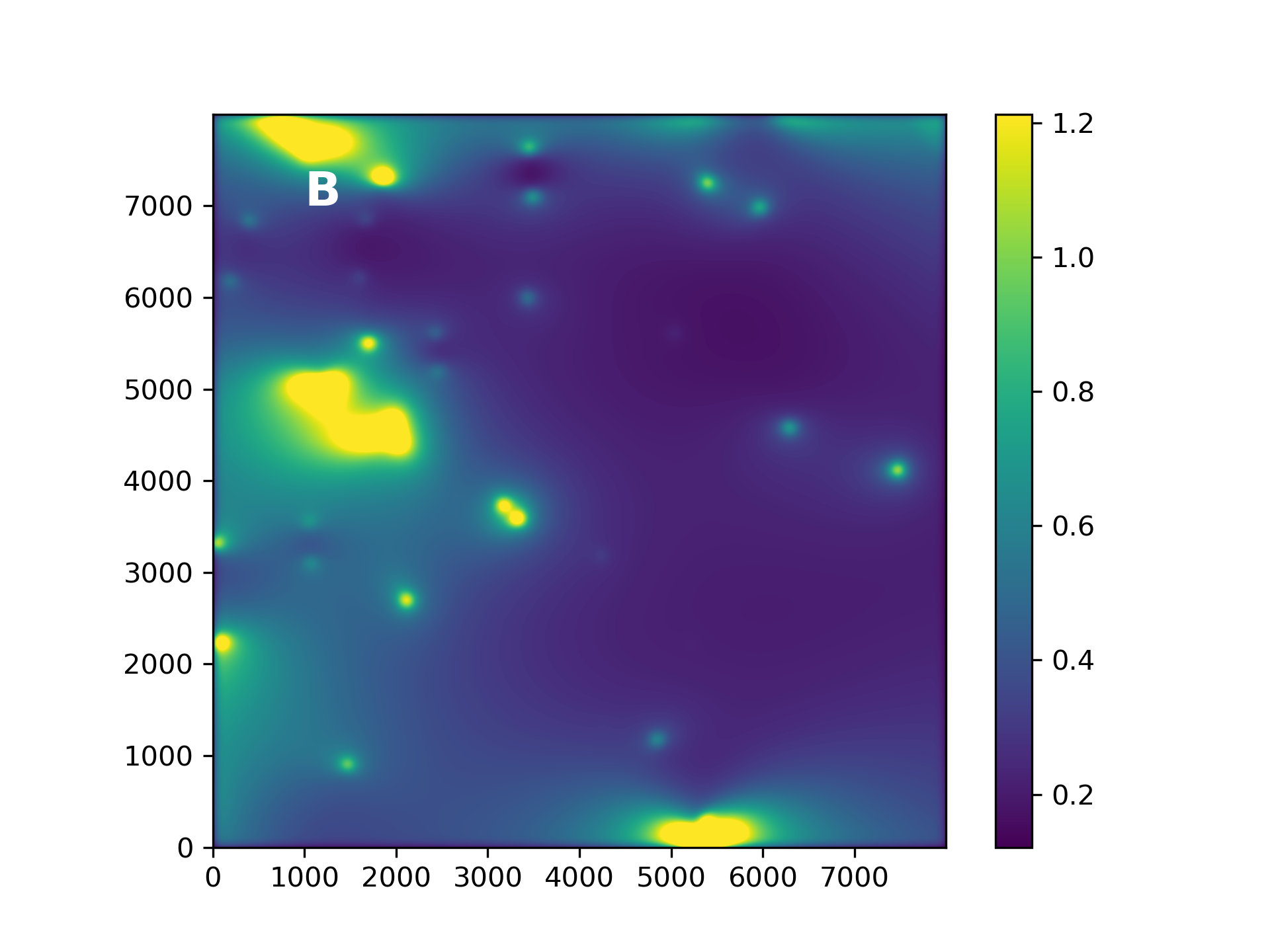}
\caption{Examples of magnification maps for the first approach. {\it First row from the 
top}: standard scenario without PBHs and with 90\% of mass in SDM. {\it Second row}: 
non-standard scenario with 50\% of mass in SDM and 45\% of mass in PBHs having $M_{\rm pbh} 
\sim$ 0.001 $\rm{M_{\odot}}$. {\it Third row}: non-standard scenario with 10\% of mass in SDM 
and 81\% of mass in PBHs having $M_{\rm pbh} \sim$ 0.1 $\rm{M_{\odot}}$. {\it Fourth row}: 
non-standard scenario with 50\% of mass in SDM and 45\% of mass in PBHs having $M_{\rm pbh} 
\sim$ 10 $\rm{M_{\odot}}$. In all scenarios, the quasar source has the intermediate size 
(see main text). The colour scale represents magnification values, with 2.66 and 0.56 being 
the macro-magnifications of A and B, respectively.}
\label{fig:examaps}
\end{figure*} 

We built square magnification patterns of 40 $R_{\rm E}$ on a side, where $R_{\rm E}$ = 
3.76$\times$10$^{16}$ cm is the Einstein radius in the source plane of a 0.3 
$\rm{M_{\odot}}$ star (for a flat $\Lambda$CDM cosmology; see Sect.~\ref{sec:lensmass}). 
These magnification maps contain $8000 \times 8000$ pixels. However, since the quasar source 
responsible for the observed $r$-band fluxes has a finite size, each magnification pattern 
was convolved with a Gaussian brightness profile $I(R) \propto \exp(-R^2/2R_{\rm s}^2)$, 
where $R_{\rm s}$ is the source radius. In this work, we considered the two values of 
$R_{\rm s}$ that were derived from previous fits of $r$-band and $H$-band/$r$-band 
microlensing variations \citep{2023ApJ...952...54R}, as well as a smaller value for 
comparison purposes (to better check the influence of the source size). Hence, by defining 
the relative radius as $r_{\rm s} = R_{\rm s}/R_{\rm E}$, we used $r_{\rm s}$ = 0.605 (fit 
of $r$-band data), 0.276 (joint fit of $H$-band and $r$-band data), and 0.1 \citep[about 15 
Schwarzschild radii for an 8.9$\times$10$^8$ $\rm{M_{\odot}}$ supermassive black hole;  
e.g.,][]{2006ApJ...649..616P}. To prevent edge
effects (Gaussian convolution biases), pixels near the sides of the convolved maps were not 
considered for subsequent analysis. The unbiased map regions contain $7200 \times 7200$ 
pixels (36 $R_{\rm E}$ on a side) and allow us to make large numbers of synthetic light 
curves with an appropriate resolution (see Sect.~\ref{sec:simmicro}).

As a summary, using the parameter values in Table~\ref{tab:massol} and those detailed in 
this section, for each approach and quasar image, we generated a total of 90 magnification 
maps, of which 9 are associated with standard scenarios without PBHs (3 SDM mass fractions 
$\times$ 3 source sizes) and 81 correspond to non-standard scenarios (3 SDM mass fractions 
$\times$ 3 PBHs mass to microlenses mass ratios $\times$ 3 PBH masses $\times$ 3 source 
sizes). The combinations of parameters that produce the total of 90 physical scenarios  
are illustrated in Table~\ref{tab:physcen}. For some scenarios, we have verified that one 
large map per approach and image is 
sufficient, as another large map from a different spatial distribution of microlenses 
produces similar results. In Figure~\ref{fig:examaps}, we show map examples for both images 
using the first approach. The top panels include magnification patterns for a standard 
scenario with 10\% of mass in stars ($f_{\rm sdmG}$ = 0.9), and the other panels 
display maps for non-standard scenarios consisting of: 45\% of mass in Jupiter-mass PBHs 
($f_{\rm sdmG}$ = 0.5; second row), 81\% of mass in PBHs with a mass similar to those of 
low-mass red dwarfs and high-mass brown dwarfs ($f_{\rm sdmG}$ = 0.1; third row), and 45\% 
of mass in PBHs with $M_{\rm pbh} \sim$ 10 $\rm{M_{\odot}}$ ($f_{\rm sdmG}$ = 0.5; fourth 
row). In the four scenarios, the intermediate size source ($r_{\rm s}$ = 0.276) is 
considered. 

\subsection{Simulated difference light curves}
\label{sec:simmicro} 

For a given image, we note that the shear direction forms an angle $\theta_{\gamma}$ with 
the celestial north, so its magnification maps are constructed using a 2D coordinate system 
in which one of the two axes coincides with this privileged (shear) direction. Therefore, 
the coordinate axes for A and B do not match, and sometimes the map of one of the two images 
is conveniently rotated to analyse source trajectories across the sky on the maps of both 
images \citep[e.g., see Fig. 2 of][]{2022A&A...659A..21P}, whereas other times the source 
trajectories on one of the two maps are rotated (our procedure; see below). For each 
scenario (set of values of $f_{\rm sdmG}$, $F_{\rm pbhG}$, $r_{\rm pbh}$, and $r_{\rm s}$) 
in the two approaches, the magnification patterns for images A and B were used to draw 
source trajectories and obtain their associated SDLCs. In order to have enough sample size, 
we generated 10$^5$ synthetic difference records for each pair of AB maps (see 
Sect.~\ref{sec:tests}).  

\begin{figure*}
\centering
\includegraphics[width=7.5cm]{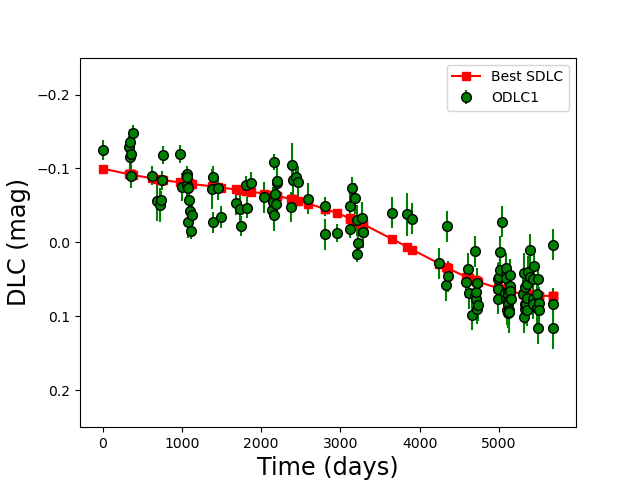}
\includegraphics[width=7.5cm]{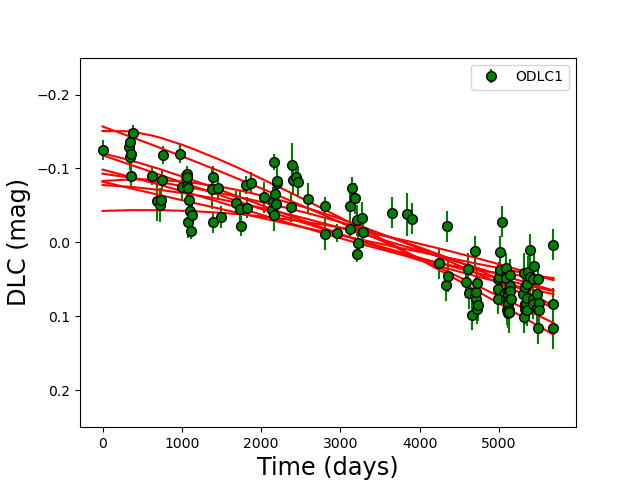}
\includegraphics[width=7.5cm]{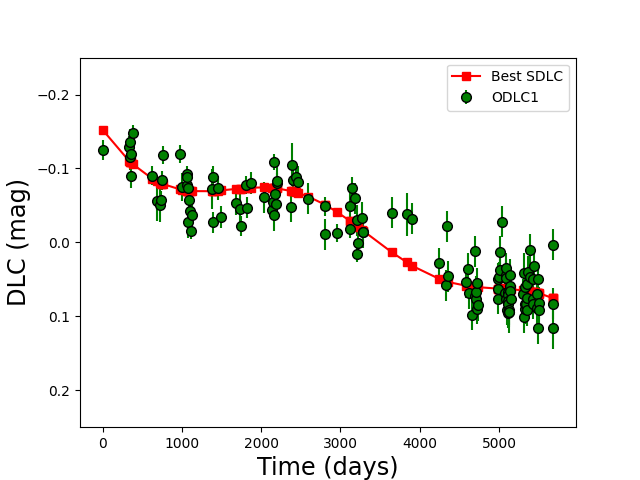}
\includegraphics[width=7.5cm]{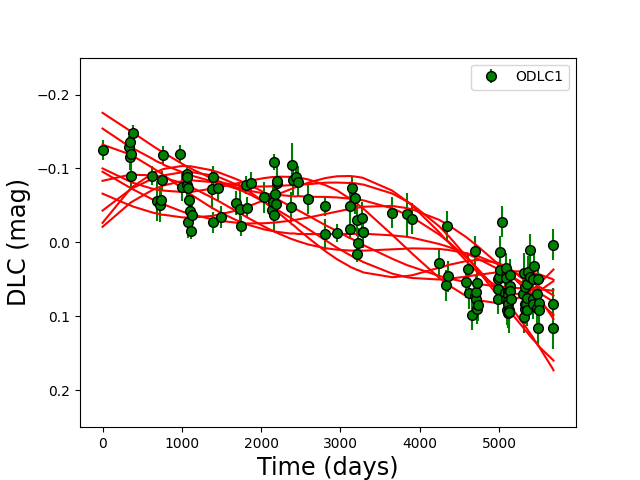}
\includegraphics[width=7.5cm]{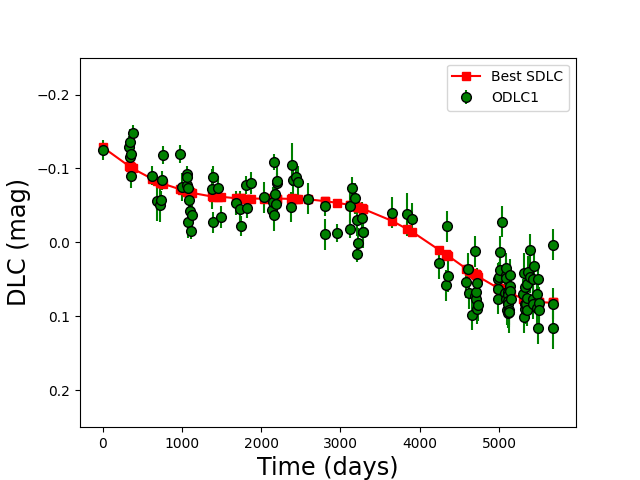}
\includegraphics[width=7.5cm]{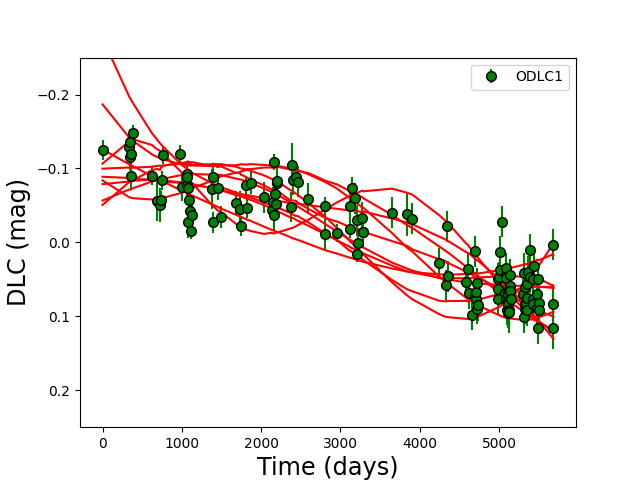}
\includegraphics[width=7.5cm]{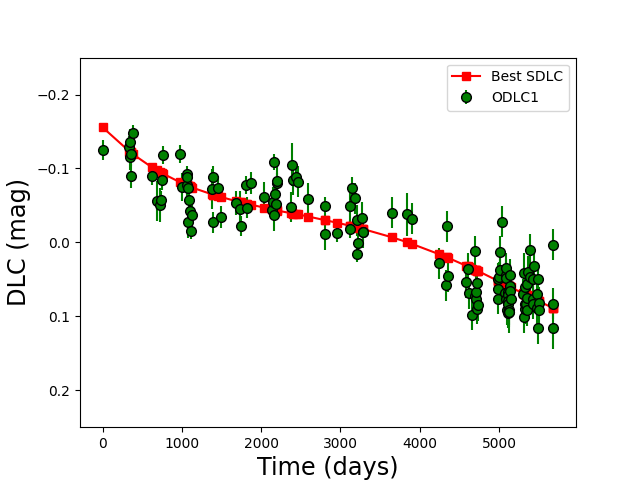}
\includegraphics[width=7.5cm]{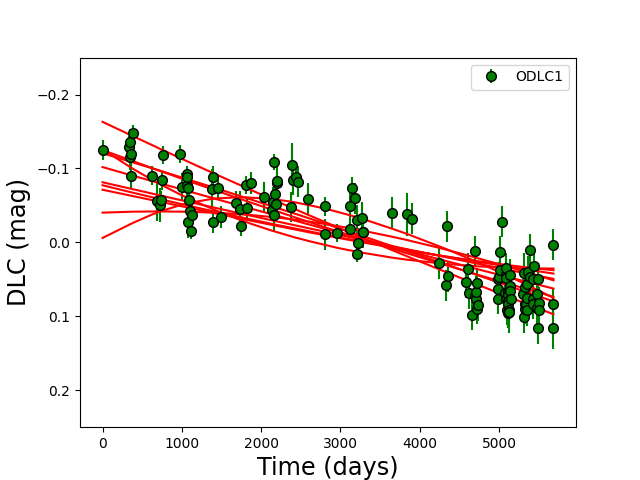}
\caption{Comparison between the ODLC in the first approach and SDLCs for all scenarios 
in Figure~\ref{fig:examaps}. After generating 10$^5$ SDLCs from each pair of maps, we show 
the best-fit SDLC (minimum $RMS$; left panels) and 10 randomly chosen SDLCs that are 
characterised by $RMS <$ 2.60 (out of a total of $n$; right panels). Each row corresponds to 
the same row position in Figure~\ref{fig:examaps}, so the results for the standard 
scenario without PBHs are depicted in the first row from the top ($RMS_{\rm min}$ = 1.67 and 
$n$ = 10\,919) and those for the non-standard scenarios are shown in successive rows: 45\% 
of mass in Jupiter-mass PBHs ($RMS_{\rm min}$ = 1.60 and $n$ = 1\,536; second row), 81\% of 
mass in $\sim$0.1 $\rm{M_{\odot}}$ PBHs ($RMS_{\rm min}$ = 1.68 and $n$ = 152; third row), 
and 45\% of mass in $\sim$10 $\rm{M_{\odot}}$ PBHs ($RMS_{\rm min}$ = 1.69 and $n$ = 
13\,381; fourth row).}
\label{fig:exasdlcs}
\end{figure*}

We estimated the effective transverse velocity of the source ($v_{\rm e}$) from Eq. (5) of 
\citet{2011ApJ...738...96M}. This effective motion provides crucial information in the time 
domain, since it links the length of a path travelled by the source (in the source plane) to 
the time elapsed in the observer's rest frame. Additionally, it depends on the redshifts of
the main deflector and source, the cosmological model, and the transverse peculiar 
velocities of the observer, main deflector, and source, as well as on the velocity 
dispersion of the microlenses in G. Using the redshifts and cosmology given in 
Sect.~\ref{sec:q0951}, and the peculiar motions and microlens velocity dispersion taken from 
\citet{2023ApJ...952...54R}, we inferred a value of $v_{\rm e}$ = 8.94$\times$10$^{7}$ cm 
s$^{-1}$. It is important to mention that $v_{\rm e}$ is basically due (95\% of the total) 
to the motion of G and microlenses within G, so its direction is unknown, and thus we can 
assume source trajectories with arbitrary orientations.

Regarding the construction of a SDLC for a source trajectory on a pair of AB maps, we took a 
random point ($x_1$, $y_1$) on the map for A and generated a random trajectory angle 
(orientation) $\alpha$ in the interval [0, 2$\pi$]. The straight path of the source on this 
A map can then be mathematically described as
\begin{equation}
\begin{split}
   x(t) &= x_1 + R(t) \cos \alpha \\
   y(t) &= y_1 + R(t) \sin \alpha ,
\end{split}
\label{eq3}
\end{equation}
where ($x_1$, $y_1$) represents the initial source position ($t$ = 0) and $R(t) = 0.041 
\times t$ is the path length (in pixels) travelled after a time $t$ (in days). We calculated 
the time differences $t_k$ = MJD$_k$ $-$ MJD$_1$ ($k = 1,...,N$) for the $N$ epochs in the 
corresponding ODLC (see Figure~\ref{fig:odlcs}), which allowed us to construct a 
magnification curve $\mu_{\rm A}(t_k)$ ($k = 1,...,N$) mimicking the ODLC sampling. For a given 
time $t_k$, the source is located at ($x_k$, $y_k$), with $x_k = x(t_k)$ and $y_k = y(t_k)$.
However, we note that $x_k$ and $y_k$ are real numbers, whereas the A map contains 
information in pixels (pair of natural numbers). Hence, we performed a weighted 
interpolation to derive the magnification at $t_k$: $\mu_{\rm A}(x_k, y_k) = [\sum_p 
\mu_{\rm A}(p) W(p)] / \sum_p W(p)$, where $\mu_{\rm A}(p)$ is the magnification in the 
pixel $p$, $W(p) = 1 - d_{pk}$ is the weight of this pixel, and $d_{pk}$ is the distance 
between the pixel $p$ and the point of interest (only pixels at distances $d_{pk} \leq 1$ 
are considered). In addition, equations for the straight path of the source on the B map are
similar to those in Eq.~\ref{eq3}, with an angle $\alpha - (\theta_{\gamma_{\rm B}} - 
\theta_{\gamma_{\rm A}})$ and another initial position, i.e., a starting point randomly 
chosen on the map for B and a path that is rotated 13\fdg6 clockwise. After computing the
corresponding magnifications $\mu_{\rm B}(t_k)$ ($k = 1,...,N$) by following the procedure 
described above, the expression on the right side of Eq.~\ref{eq2} led to the SDLC. 

\subsection{Comparison between ODLCs and SDLCs}
\label{sec:tests}

For a given scenario (in the first or second approach), we have to compare each SDLC with 
the corresponding ODLC. Here, the root mean square of relative residuals ($RMS$ for short) 
was used to assess how closely a synthetic microlensing signal matches the observed one. 
Since both signals consist of $N$ data points, the $RMS$ is given by the equation  
\begin{equation}
   RMS = \sqrt{\frac{1}{N} \sum_{j=1}^{N} \left( \frac{O_j - S_j}{E_j} \right)^2} ,
\label{eq4}
\end{equation}
where $O_j$ and $E_j$ denote the observed differences and their errors, respectively, and 
$S_j$ are the simulated values. If $RMS \sim$ 1, the SDLC fits the ODLC well, in other 
words, the SDLC is consistent with the observed data. However, the $E_j$ values could be 
slightly underestimated, leading to relatively high $RMS$ values. Thus, in a first analysis 
scheme, in order to account for a slight underestimation of errors and the presence of a few 
outliers, the ODLC-SDLC consistency threshold was set to 1.5. This means that a SDLC and the 
corresponding ODLC are consistent with each other when $RMS <$ 1.5. For example, if $RMS$ = 
1.5, the measured uncertainties are tipically 2/3 of real errors.

\begin{figure*}
\centering
\includegraphics[width=9cm]{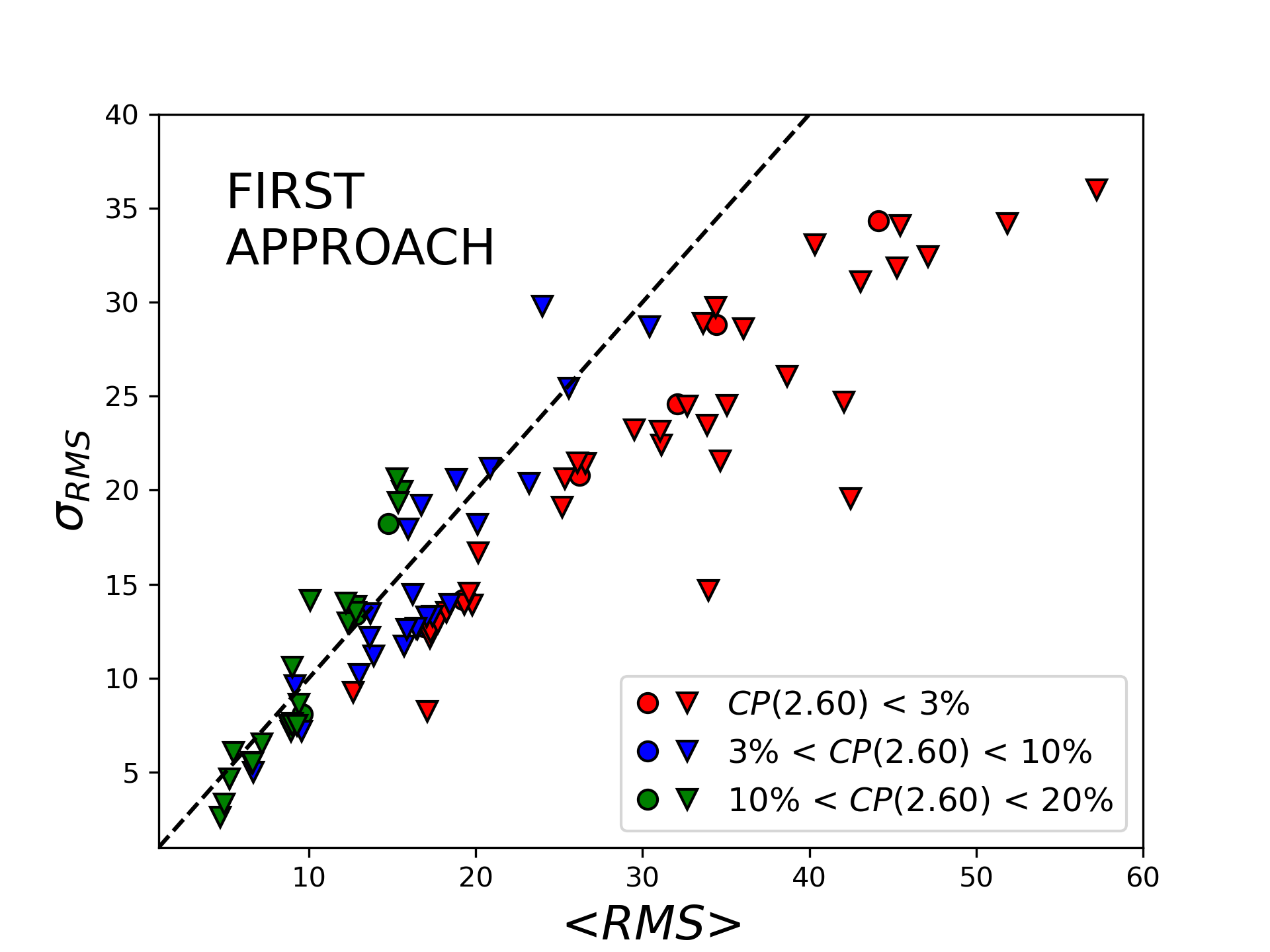}
\includegraphics[width=9cm]{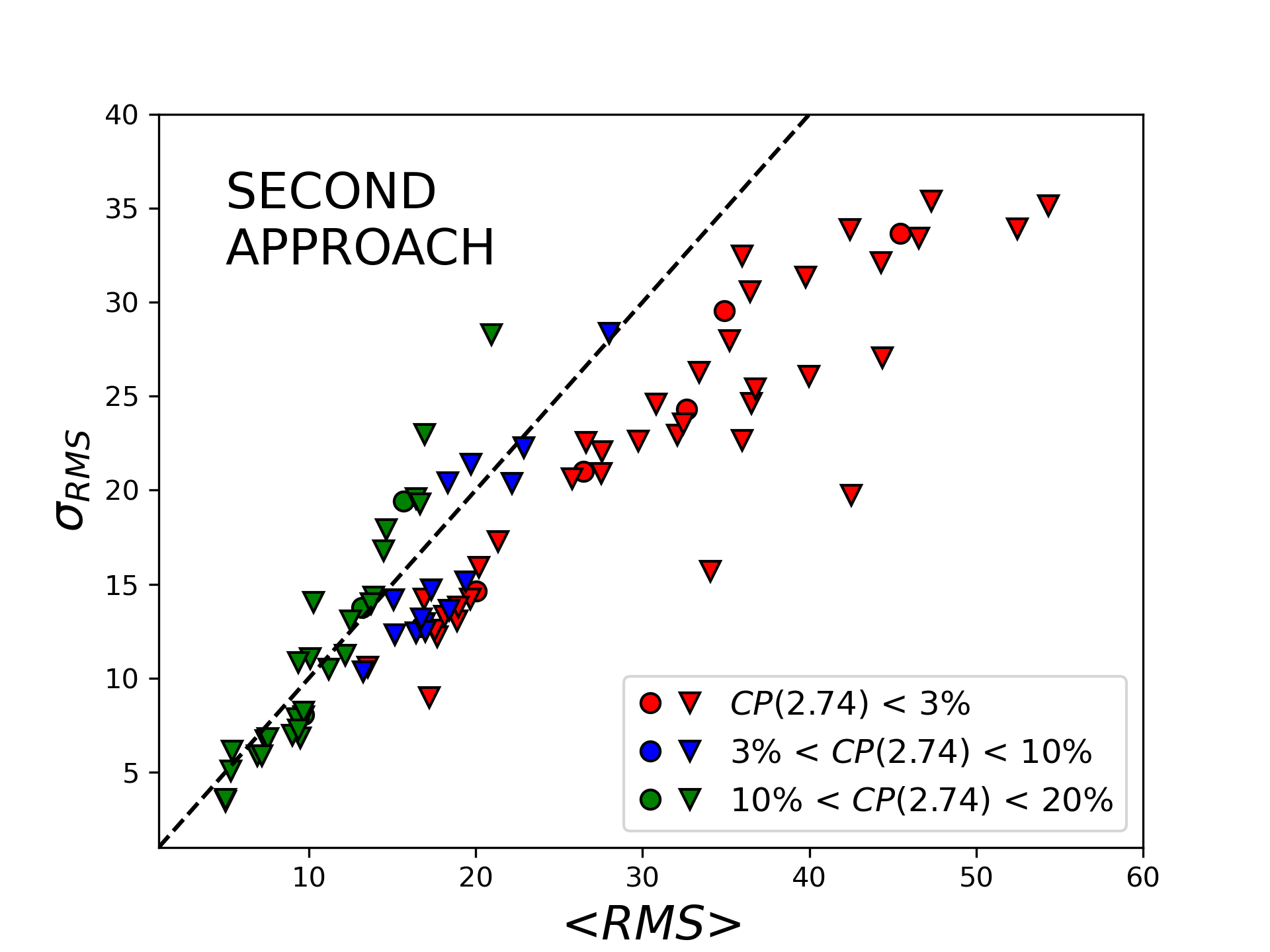}
\caption{Statistical properties of the $RMS$ distributions. For each approach in 
Table~\ref{tab:massol}, we display the $\langle RMS \rangle$ and $\sigma_{RMS}$ values 
associated with 9 standard (circles) and 81 non-standard (triangles) physical scenarios. 
Three intervals of consistency probability are also highlighted with colours red, blue, and 
green (see main text).}
\label{fig:rmstat}
\end{figure*} 

Linear fits to the ODLCs (see Figure~\ref{fig:odlcs}) trace quite well their slow 
variability. However, rapid variations also appear around these fits. Due to that, in a 
second analysis scheme, we assumed that the rapid variability in the ODLCs of \object{FBQ 
0951+2635} is not caused by standard microlensing, and consequently, our microlensing 
simulations cannot account for it. There are two plausible reasons to adopt this second 
scheme. First, observed rapid variations might be due to unaccounted, unknown observational 
noise. Second, the observed rapid variability may be due to physics that was ignored 
when constructing SDLCs, and it could be correlated with the rapid intrinsic variability 
\citep[see the last paragraph in Sect.~\ref{sec:obsmicro} and][]{2024MNRAS.530.2273G}.   
In both cases, simulations would only reproduce the long time-scale (slow) variability of 
the ODLCs, and rapid variations should be considered as additional noise. 

Therefore, for each approach given in Table~\ref{tab:massol}, original photometric errors 
were increased by a factor $RMS$(L), where $RMS$(L) is the $RMS$ value for the linear fit to 
the ODLC, while we kept the consistency threshold of 1.5 to allow for a true slow signal 
slightly different to low-order polynomial functions that fit best the data. This 
is equivalent to using a ODLC-SDLC consistency threshold of 1 and increasing the errors by a 
factor 1.5$\times$$RMS$(L), and also to using the original errors, but increasing the 
consistency thresholds up to 1.5$\times$$RMS$(L) = 2.60 and 2.74 for the first and second 
approach, respectively. To illustrate the ODLC-SDLC comparison in our second analysis 
scheme, Figure~\ref{fig:exasdlcs} displays the ODLC in the first approach along with 
well-fitted SDLCs ($RMS <$ 2.60) from the eight example maps shown in 
Figure~\ref{fig:examaps}. The minimum $RMS$ values hardly depend on the example scenario, 
and they are above 1.5 and close to $RMS$(L) $\sim$ 1.7. From a statistical point of view, 
the second and third example scenarios have more difficulties than the other two in 
reproducing the underlying long-term variation, since only $10^2-10^3$ out of 10$^5$ SDLCs 
fit the ODLC well. 

For each scenario in both approaches, we generated $N_{\rm s}$ SDLCs, $n(RMS < T)$ of 
which produce a $RMS$ value less than a threshold $T$. Thus, the consistency probability in 
\% for a threshold $T$ was defined as  
\begin{equation}
   CP(T) = 100 \times \frac{n(RMS < T)}{N_{\rm s}} ,
\label{eq5}
\end{equation}
where we considered $T$ = 1.5 (first analysis scheme), and $T$ = 2.60 or 2.74 (depending on 
the approach used; second analysis scheme). A pending issue is to discuss the choice of 
$N_{\rm s}$ (sample size), keeping in mind that a large number of possible source's paths is 
required to fully cover magnification maps and yield robust parameter estimation of $RMS$ 
distributions. After some tests, $N_{\rm s}$ = 10$^5$ provides a good compromise between 
robustness and computational cost, and Table~\ref{tab:sample} shows results from the two 
example maps in the top panels of Figure~\ref{fig:examaps} (see also top panels of 
Figure~\ref{fig:exasdlcs}). Table~\ref{tab:sample} includes uncertainties in three 
statistical parameters of the $RMS$ distribution from a sample of 10$^5$ SDLCs: mean 
($\langle RMS \rangle$), standard deviation ($\sigma_{RMS}$), and $CP$(2.60). These 
uncertainties are standard deviations (second row) derived from 10 samples of 10$^5$ SDLCs. 
It is worthy to note that the uncertainty in $CP$(2.60) is only 0.1\%. Additionally, we do 
not incorporate $CP$(1.5) into Table~\ref{tab:sample} because the consistency probability 
for a threshold $T$ = 1.5 is zero.   

\begin{table}[h!]
\begin{center}
\caption{Parameter uncertainties for a sample of size $N_{\rm s}$ = 10$^5$.}
\label{tab:sample}
\begin{tabular}{ccc}
   \hline \hline
   $\langle RMS \rangle$ & $\sigma_{RMS}$ & $CP$(2.60)\\
   \hline
   $\pm$0.027 & $\pm$0.045 & $\pm$0.095\%\\ 
   \hline
\end{tabular}
\end{center}
\footnotesize{Note: Uncertainties in the mean and standard deviation of $RMS$ values (first 
two columns), and in the consistency probability for a $RMS$ threshold of 2.60 (last column). 
See main text for details about the approach, scenario, and procedure to estimate 
uncertainties.}
\end{table}

\section{Results}
\label{sec:simres} 

Using the two approaches in Table~\ref{tab:massol}, we did not find any scenario yielding 
SDLCs with $RMS <$ 1.5. Therefore, $CP$(1.5) = 0 for the 90 physical scenarios considered in 
both approaches, covering a wide range of mass compositions of the galaxy, PBH masses, and 
source sizes. We have to emphasize here that our goal is not to achieve good fits of the 
observed variability in its entirety by fine tuning of microlensing parameters, but to use a 
limited number of microlensing parameter combinations and test how the associated scenarios 
perform. Although about twenty million (2 $\times$ 90 $\times$ 10$^5$) SDLCs failed to 
reproduce the overall variability of the ODLCs, several scenarios were reasonably consistent 
with their long time-scale variability. Figure~\ref{fig:rmstat} shows our results in the 
($\langle RMS \rangle$, $\sigma_{RMS}$) plane. The standard (circles) and non-standard 
(triangles) scenarios are classified into three categories according to their consistency 
with observations for the $RMS$ thresholds described in the last paragraph of 
Sect.~\ref{sec:tests}. Thus, the green circles/triangles correspond to the best scenarios 
with a consistency probability in the 10$-$20\% interval. In Figure~\ref{fig:rmstat}, as 
expected, these green circles/triangles are placed around or slightly above the dashed line 
$\sigma_{RMS} = \langle RMS \rangle$. We also note that a number of non-standard scenarios 
(including a PBH population) are among the best. 

\begin{figure*}
\centering
\includegraphics[width=\textwidth]{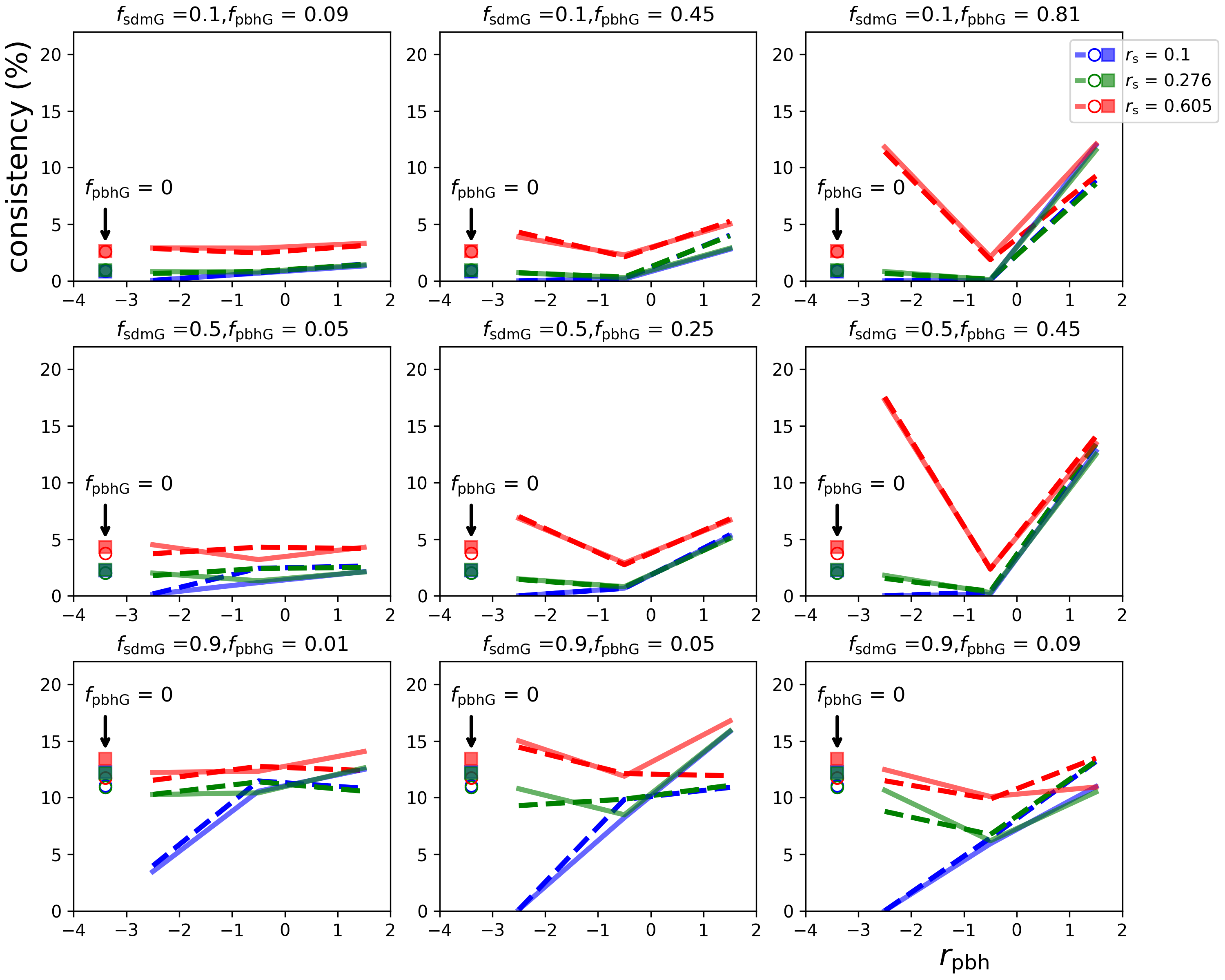}
\caption{Consistency probability between the observed slow extrinsic variability and 
simulated microlensing signals in relevant physical scenarios. The open circles (standard 
scenarios) and dashed lines (non-standard scenarios) represent the results from the first 
approach, and the filled squares (standard scenarios) and solid lines (non-standard 
scenarios) indicate the results using the second approach. The x-axis represents the 
logarithm of the $M_{\rm pbh}/M_{\rm star}$ ratio and includes the three values used in 
the non-standard scenarios, i.e., $r_{\rm pbh}$ = $-$2.5, $-$0.5, and 1.5. We have 
arbitrarily put the results for the standard scenarios at $r_{\rm pbh}$ = $-$3.4 for a 
better visual comparison. The source size is highlighted with colours blue, green, and red 
(see main text).}
\label{fig:cpv}
\end{figure*} 

All consistency probability values from our second analysis scheme are displayed in 
Figure~\ref{fig:cpv}. The nine panels of Figure~\ref{fig:cpv} are sorted so that the SDM 
mass fraction increases from top to bottom, and for a given value of $f_{\rm sdmG}$, the 
mass fraction in PBHs for non-standard scenarios increases from left to right. The open 
circles (first approach) and filled squares (second approach) denote probabilities for the 
standard scenarios with $f_{\rm pbhG}$ = 0, while the dashed (first approach) and solid 
(second approach) lines describe probabilities for the non-standard scenarios. Examining 
Figure~\ref{fig:cpv} in detail, there is a clear analogy between the results of the two 
approaches, indicating that the time delay and mass model do not play a critical role in our 
microlensing analysis. Additionally, the consistency probability is below 3\% in standard 
scenarios with $f_{\rm sdmG}$ = 0.1, and it only reaches values above 10\% when the SDM is 
the dominant contribution to the mass of the galaxy. It is also worth noting that 
microlensing effects of PBHs with a mass close to the mean stellar mass ($r_{\rm pbh}$ = 
$-$0.5 and $M_{\rm pbh} \sim$ 0.1 $\rm{M_{\odot}}$) resemble to a large extent those of 
stars, so scenarios with $f_{\rm sdmG}$ = 0.9 and a small mass fraction in PBHs ($f_{\rm 
pbhG} <$ 0.1) in this PBH mass window are favoured. These results agree with the statistical 
analysis of microlensing magnifications for a sample of 20 lensed quasars \citep[including 
\object{FBQ 0951+2635};][]{2009ApJ...706.1451M}, and the conclusions of 
\citet{2023ApJ...954..172E} about star-like PBHs.
 
In Figure~\ref{fig:cpv}, if we look at scenarios with the less massive PBHs ($r_{\rm pbh}$ = 
$-$2.5 and $M_{\rm pbh} \sim$ 0.001 $\rm{M_{\odot}}$) and the largest source, it is 
difficult to distinguish between a population of PBHs and an equivalent amount of mass in 
SDM, since PBHs behave as a smooth distribution of dark matter. For example, mass 
compositions with $f_{\rm sdmG}$ = 0.1 and $f_{\rm pbhG}$ = 0.45 (top middle panel), $f_{\rm 
sdmG}$ = 0.1 and $f_{\rm pbhG}$ = 0.81 (top right panel), and $f_{\rm sdmG}$ = 0.5 and 
$f_{\rm pbhG}$ = 0.45 (middle right panel) have probabilities similar to mass compositions 
with $f_{\rm sdmG}$ = 0.5 and $f_{\rm pbhG}$ = 0.05 (middle left panel), $f_{\rm sdmG}$ = 
0.9 and $f_{\rm pbhG}$ = 0.01 (bottom left panel), and $f_{\rm sdmG}$ = 0.9 and $f_{\rm 
pbhG}$ = 0.05 (bottom middle panel), respectively. For mass compositions of G verifying 
$f_{\rm sdmG} + f_{\rm pbhG} >$ 0.9, the probabilities vary from 11.4\% to 17.5\%, while 
they are only about 3\% for $f_{\rm sdmG} + f_{\rm pbhG} <$ 0.2. In terms of consistency, it 
is not surprising that the best scenario out of the 90 studied is the one with the largest 
source and a mass fraction in Jupiter-mass PBHs of 0.45 (see the middle right panel). This
means an "effective" SDM mass fraction of 0.95 and $f_{\rm starG}$ = 0.05, in good agreement
with Fig. 7 of \citet{2009ApJ...706.1451M}. If we focus on the smallest source, populations 
of Jupiter-mass PBHs have very small probabilities, mostly below 0.1\%. Hence, the 
constraints on the less massive PBHs strongly depend on the source size.

In addition, based on measurements of microlensing magnifications for several lensed quasars, 
previous studies indicated that $f_{\rm pbhG} \leq$ 0.01 at the 90\% confidence level for 
PBHs with 10 $\rm{M_{\odot}}$ and source sizes similar to those used in this paper 
\citep[e.g.,][]{2022ApJ...929..123E}. However, the observed slow extrinsic variability of 
\object{FBQ 0951+2635} is reasonably consistent with scenarios involving small, moderate, 
and large mass fractions in massive PBHs ($r_{\rm pbh}$ = 1.5 and $M_{\rm pbh} \sim$ 10 
$\rm{M_{\odot}}$), regardless of the source size (see the bottom and right panels of 
Figure~\ref{fig:cpv}). 

For completeness, using our second analysis scheme, in Appendix~\ref{sec:bayesanal} we also 
present results of a likelihood-based Bayesian metric for both approaches. To calculate the 
probabilities of the 90 scenarios, their individual SDLCs were weighed by $\exp(-\chi^2/2)$ 
\citep[e.g.,][]{2007ASPC..371...43K}. Additionally, taking the standard scenario with $f_{\rm 
sdmG}$ = 0.9 and the intermediate size source as a reference, the relative Bayesian 
probabilities ($\eta$) were compared with the relative consistency probabilities ($\epsilon$) in 
Figure~\ref{fig:metrics}. Although results show reasonable agreement between the two metrics, 
three spikes in $\eta$ are observed for both approaches (highlighted with black triangles in 
Figure~\ref{fig:metrics}) that have no counterparts in $\epsilon$. These high-probability spikes 
correspond to non-standard scenarios with $f_{\rm sdmG}$ = 0.9, $f_{\rm pbhG}$ = 0.01$-$0.09, 
Jupiter-mass PBHs, and the intermediate size source. The three non-standard scenarios with 
$f_{\rm sdmG}$ = 0.9, $f_{\rm pbhG}$ = 0.09, and massive PBHs also lead to relatively small 
values of $\eta$ for the second approach (see the black rectangle in the bottom panel of 
Figure~\ref{fig:metrics}). For a given scenario, while the Bayesian metric favours the best fits 
to data (minimun $\chi^2$ values), we promote the $CP$ metric ($\epsilon$ value) because it 
weighs equally all SDLCs with $RMS$ below the consistency threshold and thus less severely 
constrains the underlying signal.

\section{Conclusions}
\label{sec:conclu}

Since PBHs could provide a certain fraction of the mass in galaxies, several recent studies 
focused on the use of gravitational microlensing to probe possible PBH populations in local 
and non-local galaxies. Thus, planetary-mass and stellar-mass PBHs may only account for a 
few percent of dark matter in the Milky Way and the Large Magellanic Cloud 
\citep[e.g.,][]{2024Natur.632..749M}, while the analysis of single-epoch microlensing effects of 
non-local (mostly early-type) galaxies acting as main gravitational lenses of distant 
quasars led to relevant results on Jupiter-mass, star-like, and 10$-$60 $\rm{M_{\odot}}$ 
PBHs \citep[e.g.,][]{2009ApJ...706.1451M,2022ApJ...929..123E,2023ApJ...954..172E}. 
Microlensing effects on the Fe K$\alpha$ emission region of a few lensed quasars were also used 
to put strong constraints on the population of substellar PBHs \citep[mass fraction 
$\lesssim$0.01$-$0.03\%;][]{2018ApJ...853L..27D,2019ApJ...885...77B}.

Although difference light curves of lensed quasars is a relatively unexplored tool to 
constrain PBH populations in lensing galaxies, these time-domain microlensing signals could 
help reveal the composition of non-local galaxies. Very recently, 
\citet{2023A&A...673A..88A} used difference light curves of several lensed quasars to shed 
light on the mass composition of their main lensing galaxies. However, they only discussed 
two different physical scenarios: a standard scenario including SDM and stars, and an 
alternative scenario in which all the mass is in the form of compact objects with a stellar
mass function (only stars and stellar-mass PBHs, without SDM), concluding that both are 
consistent with the observed microlensing signals. Optical light curves of 
\object{Q2237+0305} are also consistent with $\sim$10\% of the bulge mass of its main 
(spiral) lens galaxy being formed by planetary-mass PBHs \citep{2024MNRAS.528.1979T}. In 
this paper, as a case study, we have considered difference light curves of the 
well-studied doubly imaged quasar \object{FBQ 0951+2635} spanning 16 years. The observed 
microlensing signals have been compared to synthetic difference light curves corresponding 
to source trajectories on simulated magnification maps, and we have thoroughly examined 90 
physical scenarios covering a range of relevant mass compositions of the main lensing 
galaxy, PBH masses, and source sizes. The standard scenarios only include SDM and stars, 
whereas the non-standard ones incorporate SDM, stars, and PBHs. 

The rapid variability in the observed microlensing signals cannot be explained by any of the 
90 scenarios considered. Even using a sample of 10$^8$ SDLCs for some scenarios, we did not 
find $RMS$ values below 1.5. We have, however, found several scenarios that are reasonably 
consistent with the observed slow microlensing variability, that is, best standard and 
non-standard scenarios with consistency probabilities (as defined in the last paragraph of 
Sect.~\ref{sec:tests} and shown in Figure~\ref{fig:cpv}) above 10\%. It is also demonstrated 
that the time delay and mass model adopted, provided they are reasonable, have little impact 
on the results. Regarding the standard scenarios, the best ones correspond to a galaxy mass 
dominated by SDM. Moreover, considering a PBH population with a mass close to the mean 
stellar mass, the best scenarios also include a dominant contribution of SDM to the galaxy 
mass and a relatively small mass fraction in PBHs ($<$10\%). These results using the 
variability of only one lensed quasar are in good agreement with some previous single-epoch 
studies from tens of lens systems \citep[e.g.,][]{2009ApJ...706.1451M,2017ApJ...836L..18M}.

Additionally, for the largest source, the best non-standard scenarios with Jupiter-mass PBHs 
correspond to a galaxy mass dominated by the contribution of the two non-stellar 
ingredients (SDM + PBHs), since these PBHs behave as SDM. However, for the smallest source, most 
scenarios with Jupiter-mass PBHs are highly unlikely, so the estimation of the source size
through an independent experiment \citep[e.g., using the continuum reverberation 
mapping method;][]{2012ApJ...744...47G,2018ApJ...862..123M} could lead to strong constraints on 
the population of planetary-mass PBHs in the lensing galaxy. At the other 
end of the PBH mass spectrum, some scenarios with massive PBHs ($\sim$10 $\rm{M_{\odot}}$) 
have a consistency above 10\%. For these best scenarios, the mass fraction in PBHs can even 
be $\sim$80\%. This result does not contradict the strong constraints on 10 $\rm{M_{\odot}}$ 
PBHs in a previous analysis of single-epoch fluxes of a sample of lensed quasars 
\citep[e.g.,][]{2022ApJ...929..123E}, but rather encourages the study of light
curves with longer temporal coverage to try to confirm/improve current constraints by 
microlensing variability. Our ODLCs describe microlensing variations on a spatial scale
similar to the Einstein radius in the source plane of a 0.3 $\rm{M_{\odot}}$ star, which is
equivalent to about 20\% of the Einstein radius of a 10 $\rm{M_{\odot}}$ microlens. This justify 
the high sensitivity to objects with 0.1$-$0.3 $\rm{M_{\odot}}$ and the need of longer 
trajectories to constrain the population of massive PBHs. The ODLCs also span $\sim$20 Einstein 
radii of a Jupiter-mass microlens, making them extraordinarily sensitive to these 
planetary-mass objects for the most compact source. 

As a general conclusion, variability and single-epoch studies of lensed quasars have great 
potential to constrain PBH populations, but the data, methodology, and targets used may play 
an important role. Future work with optical light curves\footnote{It would also be desirable 
to have X-ray flux time series to constrain populations of sub-Jupiter mass PBHs} should be 
based on records with low observational noise and spanning tens of years (covering long 
trajectories in magnification maps), as well as reverberation-based source sizes and 
microlensing simulations with a fine resolution of the parameter grid (it might be required 
to use a double source, e.g., accretion disc plus broad line region). There is also the 
pending task of probing the possible presence of PBHs along the line of sight to a given lensed 
quasar \citep[e.g.,][]{2020A&A...633A.107H,2022MNRAS.512.5706H,2024MNRAS.527.2393H} from complex 
simulations involving PBH populations at different redshifts. 

\section{Data availability}
\label{sec:data}

The light curves and Python scripts to create/show ODLCs are publicly available at 
https://github.com/glendama/q0951odlc, and the Fortran-90 code to generate magnification maps is 
publicly available at https://github.com/glendama/magmaps.

\begin{acknowledgements}
The Liverpool Telescope is operated on the island of La Palma by Liverpool John Moores 
University in the Spanish Observatorio del Roque de los Muchachos of the Instituto de 
Astrof\'isica de Canarias with financial support from the UK Science and Technology 
Facilities Council. AEG acknowledges support from project ANID Fondecyt Postdoctorado 
with grant number 3230554. JMD acknowledges support from project PID2022-138896NB-C51 
(MCIU/AEI/MINECO/FEDER, UE) Ministerio de Ciencia, Investigación y Universidades. This 
research has been supported by Universidad de Cantabria funds and the grant 
PID2020-118990GB-I00 funded by MCIN/AEI/10.13039/501100011033. 

\end{acknowledgements}

\begin{appendix}

\section{Summary of the physical scenarios}
\label{sec:sumscen}

\begin{table}[h!]
\begin{center}
\caption{The 30 compositions of the lensing galaxy for each of the three source sizes: 
$r_{\rm s}$ = 0.605 (largest), 0.276 (intermediate), and 0.1 (smallest).}
\label{tab:physcen}
\begin{tabular}{ccccc}
   \hline \hline
   $f_{\rm sdmG}$ & $F_{\rm pbhG}$ & $f_{\rm starG}$ & $f_{\rm pbhG}$ & $r_{\rm pbh}$\\
   \hline 
   0.1 & 0.0 & 0.90 & 0.00 &  ---  \\
       & 0.1 & 0.81 & 0.09 & $-$2.5\\
       &     &      &      & $-$0.5\\
       &     &      &      & $+$1.5\\
       & 0.5 & 0.45 & 0.45 & $-$2.5\\
       &     &      &      & $-$0.5\\
       &     &      &      & $+$1.5\\
       & 0.9 & 0.09 & 0.81 & $-$2.5\\
       &     &      &      & $-$0.5\\
       &     &      &      & $+$1.5\\
   0.5 & 0.0 & 0.50 & 0.00 &  ---  \\
       & 0.1 & 0.45 & 0.05 & $-$2.5\\
       &     &      &      & $-$0.5\\
       &     &      &      & $+$1.5\\
       & 0.5 & 0.25 & 0.25 & $-$2.5\\
       &     &      &      & $-$0.5\\
       &     &      &      & $+$1.5\\
       & 0.9 & 0.05 & 0.45 & $-$2.5\\
       &     &      &      & $-$0.5\\
       &     &      &      & $+$1.5\\
   0.9 & 0.0 & 0.10 & 0.00 &  ---  \\
       & 0.1 & 0.09 & 0.01 & $-$2.5\\
       &     &      &      & $-$0.5\\
       &     &      &      & $+$1.5\\
       & 0.5 & 0.05 & 0.05 & $-$2.5\\
       &     &      &      & $-$0.5\\
       &     &      &      & $+$1.5\\
       & 0.9 & 0.01 & 0.09 & $-$2.5\\
       &     &      &      & $-$0.5\\
       &     &      &      & $+$1.5\\
   \hline
\end{tabular}
\end{center}
\footnotesize{Note: $f_{\rm sdmG}$, $f_{\rm starG}$, and $f_{\rm pbhG}$ are the mass 
fractions of the galaxy G in SDM, stars, and PBHs, respectively. The stellar mass 
function follows a Kroupa distribution with a mean mass $M_{\rm star}$ = 0.3 
$\rm{M_{\odot}}$, while all PBHs have the same mass $M_{\rm pbh}$, where $r_{\rm pbh} 
= \log (M_{\rm pbh}/M_{\rm star})$. The relative radius $r_{\rm s}$ is the ratio
between the source radius and the Einstein radius (in the source plane) of a typical 
star with the mean mass $M_{\rm star}$.}
\end{table}

\section{Comparing the consistency probability with a simple Bayesian metric}
\label{sec:bayesanal}

In this paper, we use a dataset ODLC to draw inferences about a physical scenario $s$. Thus, 
in a Bayesian framework for quasar microlensing \citep[e.g.,][]{2007ASPC..371...43K}, the 
probability of a pair of trajectories $\tau$ in the two magnification maps for $s$ given the 
data is
\begin{equation}
   P(s, \tau|{\rm ODLC}) \propto  P({\rm ODLC}|s, \tau) \propto \exp[-\chi^2(s, \tau)/2] ,
\label{eq6}
\end{equation}
where $P({\rm ODLC}|s, \tau)$ is the likelihood of the data given ($s$, $\tau$) and 
\begin{equation}
   \chi^2(s, \tau) = \sum_{j=1}^{N} \left[ \frac{O_j - S_j(s, \tau)}{E_j} \right]^2 .
\label{eq7}
\end{equation}
We computed the $\chi^2$ value of each SDLC for $s$, and then estimated the probability of 
the scenario $s$ by marginalising over all trajectories. This means we summed the 
probabilities for the randomly sampled trajectories to obtain
\begin{equation}
   P(s|{\rm ODLC}) \propto \sum_{\tau} \exp[-\chi^2(s, \tau)/2] .
\label{eq8}
\end{equation}

\begin{figure}[h!]
\centering
\includegraphics[width=9cm]{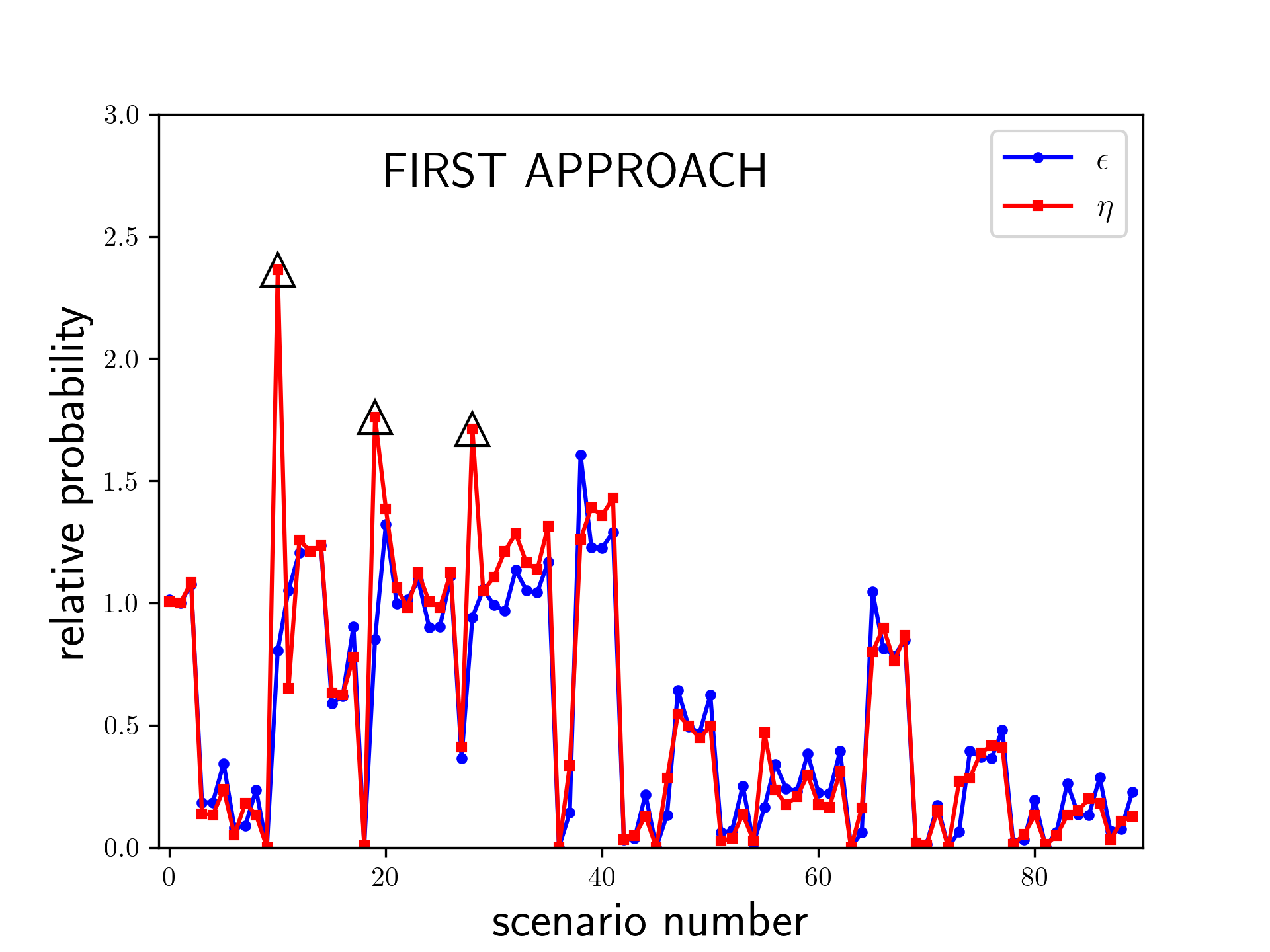}
\includegraphics[width=9cm]{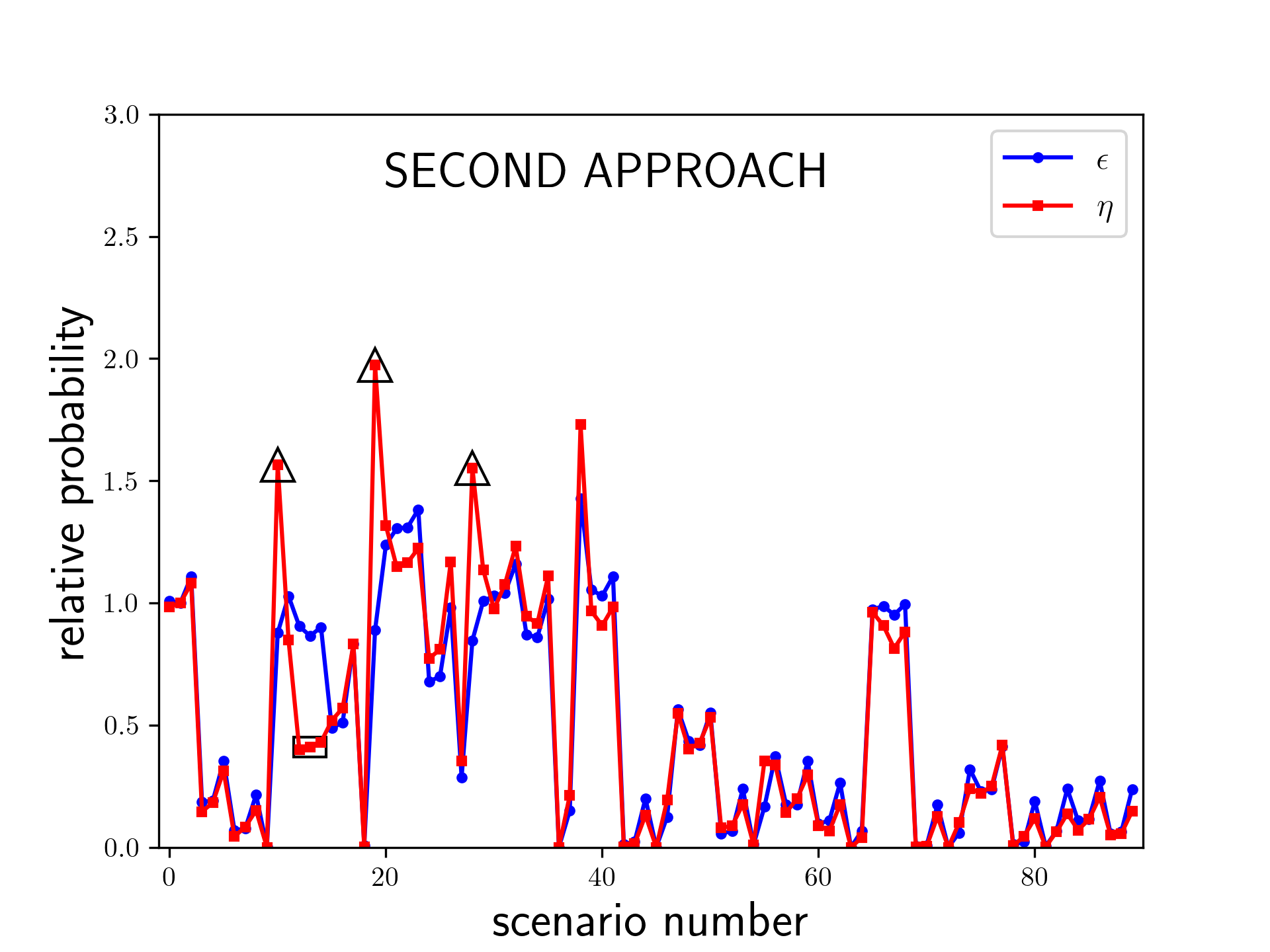}
\caption{Relative probabilities from two different metrics. Considering the two approaches 
in Table~\ref{tab:massol}, we compare the values of $\epsilon$ and $\eta$ for each of the 90
physical scenarios summarised in Appendix~\ref{sec:sumscen}. The values of $\eta$ for the
same three scenarios in both approaches are highlighted with black triangles. We also use a 
black rectangle to highlight the values of $\eta$ for three other scenarios in the second 
approach (see main text).}
\label{fig:metrics}
\end{figure}

Using formal photometric errors $E_j$, as expected, the SDLCs failed to reproduce the 
overall variability of the ODLCs, since the $\chi^2/N$ values exceeded 2.5 (first approach) 
and 2.7 (second approach). When increasing errors by a factor 1.5$\times$$RMS$(L) (see the 
second analysis scheme in Sect.~\ref{sec:tests}), we calculated probabilities relative to a 
reference scenario $s_*$: $\eta(s) = P(s|{\rm ODLC})/P(s_*|{\rm ODLC})$, where $s_*$ is the 
standard scenario consisting of a lensing galaxy with 10\% of mass in stars and the 
intermediate size source (see the top panels of Figures~\ref{fig:examaps} and 
\ref{fig:exasdlcs}). The $\eta(s)$ values were then compared with the relative probabilities 
from the $CP$ metric, i.e., $\epsilon(s) = CP(T)/CP_*(T)$, where $CP_*$ is the consistency 
probability for $s_*$ and $T$ = 1.5$\times$$RMS$(L). Figure~\ref{fig:metrics} shows our 
results for both approaches using the two metrics. These results are discussed at the end of 
Sect.~\ref{sec:simres}.

\end{appendix}

\end{document}